\documentclass[preprint]{aastex63}

\newcommand{\sm}{$M_\odot$}
\newcommand{\sL}{$L_\odot$}

\newcommand{\hcccn}{HC$_{3}$N}

\newcommand{\co}{C$^{18}$O}
\newcommand{\ccchh}{c-C$_3$H$_2$}

\newcommand{\cchoh}{C$_2$H$_5$OH}
\newcommand{\meta}{CH$_3$OH}

\newcommand{\nhhcho}{NH$_2$CHO}

\newcommand{\kms}{km s$^{-1}$}
\newcommand{\mjybeam}{mJy beam$^{-1}$}
\newcommand{\metd}{CH$_2$DOH}

\received{}
\revised{\today}
\accepted{July 11, 2022}
\submitjournal{ApJ}

\shorttitle{}
\shortauthors{Okoda et al.}

\begin{document}

\title{Chemical Differentiation and Temperature Distribution on a Few au Scale around the Protostellar Source B335}

\correspondingauthor{Yuki Okoda}
\email{yuki.okoda@riken.jp}


\author{Yuki Okoda}
\affiliation{RIKEN Cluster for Pioneering Research, 2-1, Hirosawa, Wako-shi, Saitama 351-0198, Japan}
\affiliation{Department of Physics, The University of Tokyo, 7-3-1, Hongo, Bunkyo-ku, Tokyo 113-0033, Japan}

\author{Yoko Oya}
\affiliation{Department of Physics, The University of Tokyo, 7-3-1, Hongo, Bunkyo-ku, Tokyo 113-0033, Japan}
\affiliation{Research Center for the Early Universe, The University of Tokyo, 7-3-1, Hongo, Bunkyo-ku, Tokyo 113-0033, Japan} 

\author{Muneaki Imai}
\affiliation{Department of Physics, The University of Tokyo, 7-3-1, Hongo, Bunkyo-ku, Tokyo 113-0033, Japan}

\author{Nami Sakai}
\affiliation{RIKEN Cluster for Pioneering Research, 2-1, Hirosawa, Wako-shi, Saitama 351-0198, Japan}

\author{Yoshimasa Watanabe}
\affiliation{Materials Science and Engineering, College of Engineering, Shibaura Institute of Technology, 3-7-5 Toyosu, Koto-ku, Tokyo 135-8548, Japan}

\author{Ana L\'{o}pez-Sepulcre}
\affiliation{Univ. Grenoble Alpes, CNRS, IPAG, 38000 Grenoble, France}
\affiliation{Institut de Radioastronomie Millim\'{e}trique, 38406 Saint-Martin d'H$\grave{e}$res, France}

\author{Kazuya Saigo}
\affiliation{Department of Physics, The University of Tokyo, 7-3-1, Hongo, Bunkyo-ku, Tokyo 113-0033, Japan}

\author{Satoshi Yamamoto}
\affiliation{Department of Physics, The University of Tokyo, 7-3-1, Hongo, Bunkyo-ku, Tokyo 113-0033, Japan}
\affiliation{Research Center for the Early Universe, The University of Tokyo, 7-3-1, Hongo, Bunkyo-ku, Tokyo 113-0033, Japan} 


\begin{abstract}
\par Resolving physical and chemical structures in the vicinity of a protostar is of fundamental importance for elucidating their evolution to a planetary system.
In this context, we have conducted 1.2 mm observations toward the low-mass protostellar source B335 at a resolution of 0$''$03 with ALMA.
More than 20 molecular species including HCOOH, NH$_2$CHO, HNCO, CH$_3$OH, CH$_2$DOH, CHD$_2$OH, and CH$_3$OD are detected within a few 10 au around the continuum peak.
We find a systematic chemical differentiation between oxygen-bearing and nitrogen-bearing organic molecules by using the principal component analysis for the image cube data.
The distributions of the nitrogen-bearing molecules are more compact than those of the oxygen-bearing ones except for HCOOH.
The temperature distribution of the disk/envelope system is revealed by a multi-line analysis for each of HCOOH, NH$_2$CHO, CH$_3$OH, and CH$_2$DOH.
The rotation temperatures at the radius of 0\farcs06 along the envelope direction of CH$_3$OH and CH$_2$DOH are derived to be 150-165 K.
On the other hand, those of HCOOH and NH$_2$CHO, which have a smaller distribution, are 75-112 K, and are significantly lower than those for CH$_3$OH and CH$_2$DOH.
This means that the outer envelope traced by CH$_3$OH and CH$_2$DOH is heated by additional mechanisms rather than the protostellar heating.
We here propose the accretion shock as the heating mechanism.
The chemical differentiation and the temperature structure on a few au scale provide us with key information to further understand chemical processes in protostellar sources.

\end{abstract}


\section{Introduction}\label{intro}
\par Complex organic molecules (COMs), consisting of more than six atoms with one carbon atom at least, are important species to understand the molecular evolution in the Universe.
They have extensively been found in various environments of low-mass star formation: quiescent dense clouds, prestellar cores \citep[e.g.,][]{Bacmann et al.(2012), Cernicharo et al.(2012), Ceccarelli et al.(2017), Soma et al.(2018), ScibelliShirley(2020), Jimenez-Serra et al.(2021), Scibelli et al.(2021)}, disk/envelope systems of protostellar cores \citep[e.g.,][]{Cazaux et al.(2003), Bottinelli et al.(2004)b, Kuan et al.(2004), Pineda et al.(2012), Jorgensen et al.(2016), Oya et al.(2016), Lee et al.(2019), Imai et al.(2019), Bianchi et al.(2020), Manigand et al.(2020), van Gelder et al.(2020), Martin-Domenech et al.(2021), Nazari et al.(2021), Tychoniec et al.(2021), Chahine et al.(2022)}, and outflow shock regions around protostars \citep[e.g.,][]{Arce et al.(2008), Sugimura et al.(2011), Codella et al.(2020), De Simone et al.(2020)}.
A chemical differentiation between nitrogen-bearing and oxygen-bearing species has long been recognized in high-mass protostellar clusters \citep[e.g.,][]{Blake et al.(1987), Feng et al.(2015), Wright et al.(1996), Wyrowski et al.(1999)} and a low-mass binary source \citep{Kuan et al.(2004), Caux et al.(2011)}.
In recent years, ALMA observations reveal the small-scale chemical differentiation in each single protostellar source \citep{Oya et al.(2017), Csengeri et al.(2019), Okoda et al.(2021), Nazari et al.(2021)}, where the nitrogen-bearing ones tend to have a more compact distribution than oxygen-bearing ones. 
Further differentiation among oxygen-bearing molecules has been reported.
For instance, HCOOH shows a similar trend to the nitrogen-bearing COMs in low-mass and high-mass protostellar sources, although it is classified as an oxygen-bearing molecule \citep[e.g.,][]{Oya et al.(2017), Csengeri et al.(2019)}.
The observational results indicate increasing complexity of chemical structures on a small scale around a protostar in interstellar clouds and star forming regions.
\par Although the chemical processes responsible to the chemical differentiation have been investigated in the laboratory \citep[e.g.,][]{Ioppolo et al.(2011)} and theoretical works \citep[e.g.,][]{Charnley et al.(1992), Caselli et al.(1993), Garrod et al.(2008), Vasyunin et al.(2017), Aikawa et al.(2020), Garrod et al.(2022)}, our understanding is far from complete.
According to the chemical network calculations by \cite{Aikawa et al.(2020)}, oxygen-bearing COMs and nitrogen-bearing COMs appear in the almost same region around the protostar.
Thus, the observed chemical differentiation is not well explained theoretically, and the chemical pathways/formation mechanisms responsible to the differentiation are still an open question.
To step forward, the temperature structure around the protostar needs to be clarified in detail, because it is tightly related to the chemical structure \citep[e.g.,][]{Jorgensen et al.(2018), Taquet et al.(2019), Oya(2020), van Gelder et al.(2020), van't Hoff et al.(2020), van't Hoff et al.(2020)b, Ambrose et al.(2021)}.
Hence, we here investigate molecular distributions and temperature structure around the low-mass protostellar source B335 which is rich in COMs.
\par  B335 is a Bok globule \citep{Keene et al.(1980)} harboring the Class 0 protostellar source, IRAS 19347+0727.
Since this source is in an isolated condition without influences of nearby star-formation activities, it has been used as an excellent testbed for star-formation studies \citep[e.g.,][]{Hirano et al.(1988), Evans et al.(2015), Yen et al.(2015b), Bjerkeli et al.(2019), Imai et al.(2016), Imai et al.(2019)}.
The distance to B335 was reported to be 90-120 pc, which was evaluated from stellar spectral energy distributions of nearby stars \citep{Olofsson(2009)}.
Therefore, 100 pc was often employed as the conventional distance to B335 in the previous works \citep*[e.g.,][]{Bjerkeli et al.(2019), Imai et al.(2016), Imai et al.(2019)}.
Recently, \cite{Watson(2020)} reported the distance to HD184982, which is thought to be associated with B335, to be 164.5 pc, by using the Gaia DR2 parallax data, and hence, we employ the 165 pc distance in this study.
Hereafter, we represent the physical values corrected for the distance of 165 pc, when we refer the previous results.
The bolometric temperature ($T_{\rm bol}$) and the bolometric luminosity ($L_{\rm bol}$) are 37 K \citep{Andre et al.(2000)} and 1.6 \sL\ \citep{Kang et al.(2021)}, respectively.
A bipolar outflow extending along the east-to-west direction (P.A. $\sim$90$^\circ$) has been detected with single-dish and interferometric observations \citep[e.g.,][]{Hirano et al.(1988), Hirano et al.(1992), Stutz et al.(2008), Bjerkeli et al.(2019)}.
It is almost parallel to the plane of the sky, where the inclination angle is reported to be between 3$^\circ$ and 10$^\circ$ \citep{Hirano et al.(1988), Stutz et al.(2008)}.
\par Some observational works on the disk/envelope system of this source have been carried out with ALMA.
 \cite{Evans et al.(2015)} and \cite{Yen et al.(2015b)} indicate that the infall motion dominates on a 100 au scale around the protostar without a clear rotation motion at a spatial resolution of 0.$''$3.
 \cite{Yen et al.(2015b)} estimate the protostellar mass to be 0.08 \sm\ from the marginal rotation feature seen in the \co\ data by assuming the infalling-rotating motion.
Later, higher angular resolution observations have revealed a clear rotation structure along the disk/envelope system (P.A. $\sim$0$^\circ$) at $\sim$15 au scale \citep{Imai et al.(2019)}.
The methanol (\meta) and formic acid (HCOOH) lines are seen in the vicinity of the protostar, which reveal the velocity gradient along the envelope direction.
\cite{Imai et al.(2019)} apply the infalling-rotating envelope (IRE) model to these velocity structures, and derive the protostellar mass and the radius of the centrifugal barrier to be 0.03$-$0.1 \sm\ and $<$8 au, respectively.
\cite{Bjerkeli et al.(2019)} independently report the velocity gradient in the \meta\ and SO$_2$ lines, and find that it is consistent with a pure free fall or a Keplerian rotation with the protostellar mass of 0.08 \sm.
They also estimate the dust mass from the dust continuum emission within 12 au to be 8 $\times$ 10$^{-4}$ \sm, which is comparable to the mass 2.0 $\times$ 10$^{-3}$ \sm\ derived for the area within the radius of 41 au by \cite{Evans et al.(2015)}.
\par As well as the above physical structures, B335 has interesting features from the chemical point of view. 
The CO, CN, HCO$^+$, HCN, HNC, N$_2$H$^+$, H$_2$CO, CS, CCH, and \ccchh\ lines were detected on a few 100$-$1000 au scale with ALMA \citep{Evans et al.(2015), Imai et al.(2016)}.
Distributions of the CCH and \ccchh\ emission around the protostar indicate the warm carbon-chain chemistry (WCCC) nature of this source \citep{Sakai et al.(2008), SakaiandYamamoto(2013)}.
For the inner region at a few 10 au scale, various COM lines, such as acetaldehyde (CH$_3$CHO), methyl formate (HCOOCH$_3$), and formamide (\nhhcho), were detected with ALMA by \cite{Imai et al.(2016)}.
This feature is the characteristic of a hot corino \citep{Cazaux et al.(2003), HerbstandvanDishoeck(2009)}.
Hence, B335 is recognized as the WCCC source containing a hot corino: namely a hybrid source like L483 \citep{Oya et al.(2017)} and CB68 (Imai et al 2022).
Since the chemical structure of hybrid sources is consistent with the chemical model \citep{Aikawa et al.(2008), Aikawa et al.(2020)}, they are thought to be a common occurrence.
However, their identification requires high-sensitivity and high-resolution observations, which makes a small member of the identified sources.
Note that we now know more than 20 hot corino sources and a few WCCC sources, some of which may be hybrid sources.
\par Nevertheless, the hot corino of B335 is so tiny (within a few 10 au) that it has not well been resolved spatially. 
This situation prevents us from detailed discussions of the chemical processes occurring there. 
To further elucidate physical and chemical structures of the disk/envelope system within a few 10 au scale and their mutual relation, we have observed COMs at the angular resolution (0$\farcs03\times0\farcs$02) with ALMA.
We explore chemical differentiation among COMs in the hot corino by using the Principal Component Analysis for the Cube Data (PCA-3D) in Section \ref{Chemical Differentiation}. 
We also derive the temperature structure within a hot corino from the multiline analyses of HCOOH, \nhhcho, \metd\ and  \meta\ (Section \ref{Temperature Distribution}), and discuss it in relation to the chemical differentiation (Section \ref{HCOOH}).

\section{Observation}\label{b335_observation_accretion}
\par Four blocks of the single-point ALMA observations toward B335 were carried out with the Band 6 receiver in the Cycle 6 operation on 2019 June 10, 12, 13, and 23. 
The observation parameters are summarized in Table \ref{ALMA_observations_B335}.
We combined these visibility data in the $uv$ plane.
The field center was taken to be ($\alpha_{2000}$, $\delta_{2000}$)= (19$^{\rm h}$37$^{\rm m}$00$^{\rm s}$.898, $+$07\arcdeg34\arcmin09.$''$528).
In all the observations, 52 antennas were used, where the baseline length ranged from 83 to 16196 m.
The total on-source time was 180.2 minutes.
The primary beam (half-power beam) width was 23.$''$7.
The largest angular size is 0$.''$3 for all these observations.
The bandpass calibrator, the flux calibrator, and the pointing calibrator were J1924-2914.
The phase calibrator was J1938+0448.
\par The data were reduced by Common Astronomy Software Applications package (CASA) 5.8.0 \citep{McMullin et al.(2007)} as well as a modified version of the ALMA calibration pipeline.
Phase self-calibration was performed by using the continuum data, and then the solutions were applied to the spectral line data. 
After the self-calibration procedures, the data images were prepared by using the CLEAN algorithm, where the Briggs' weighting with a robustness parameter of 0.5 was employed.
The original synthesized beam size is typically 0$\farcs03\times0\farcs$02 (P.A. -30\degr).

\section{Small-Scale Chemical Differentiation}\label{Chemical Differentiation}
\subsection{Dust Continuum}
\par
\begin{figure}[h!]
\centering

\includegraphics[scale=0.5]{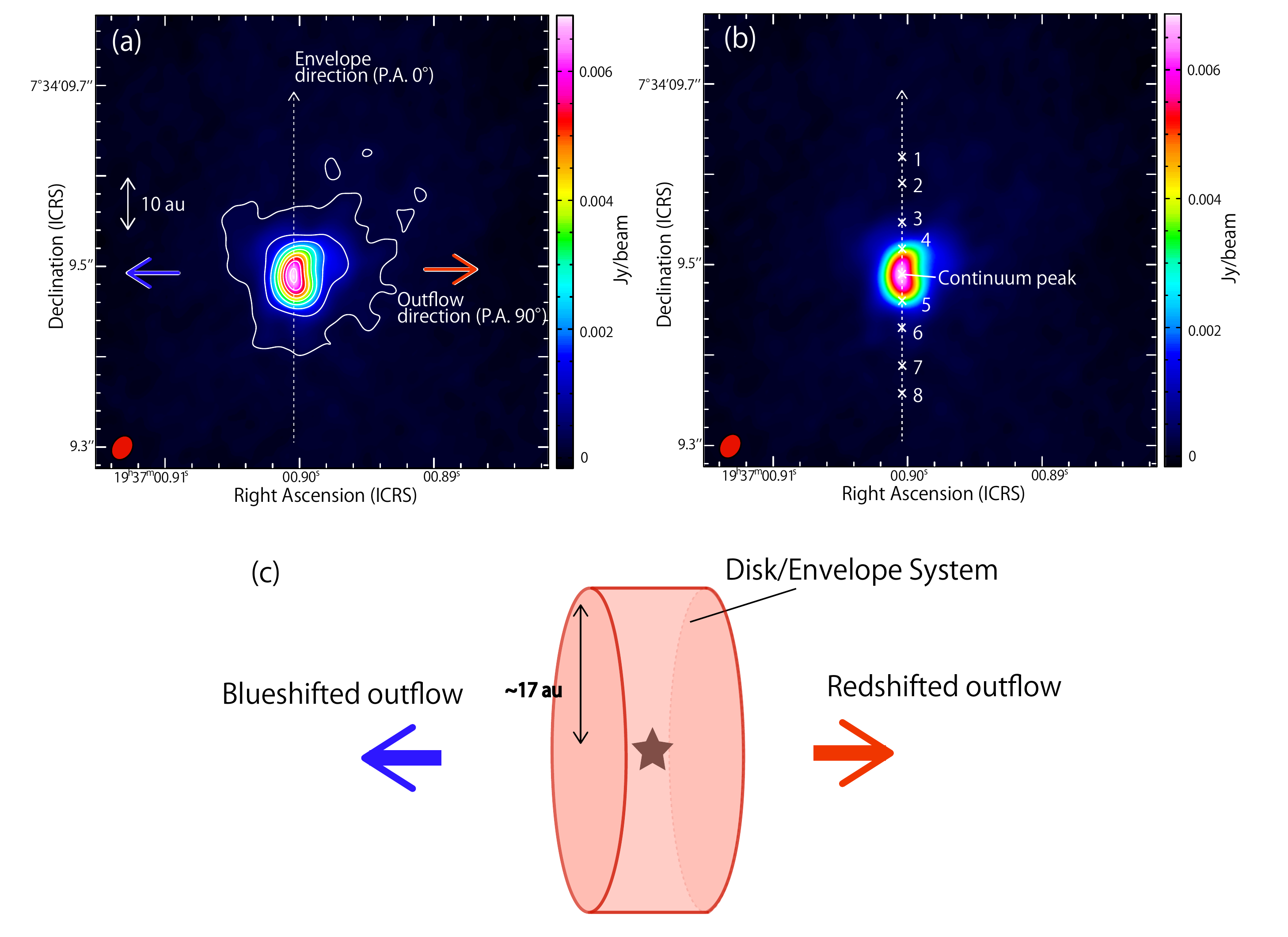}
\caption{(a,b) Dust continuum emission at 1.2 mm.
(a) Contour levels are every 10$\sigma$ from 3$\sigma$, where $\sigma$ is 0.08 \mjybeam.
The horizontal and vertical dashed arrows show the outflow and envelope directions, respectively.
(b) The cross marks in the right panel correspond to the positions for the derivation of the rotation temperature and the column density (Section \ref{temperature_distribution}; Table \ref{disktemp}).
The numbers indicate the positions presented in Table \ref{disktemp}.
(c) Schematic illustration of the disk/envelope system and outflow configuration in B335.
\label{cont_B335}}
\end{figure}

 Figures \ref{cont_B335}(a) and (b) show the dust continuum emission at 1.2 mm observed in this study.
It is single-peaked with a slight elongation along the north-south direction, which is consistent with the disk/envelope direction (P.A. $\sim$0\degr).
This feature is similar to the continuum emission reported previously \citep{Imai et al.(2016), Bjerkeli et al.(2019)}.
As well, the dust emission in this work shows a faint extension to the northwest.
Its peak position and peak intensity are derived from a 2D Gaussian fit to the image to be ($\alpha_{2000}$, $\delta_{2000}$) = (19$^{\rm h}$37$^{\rm m}$00$^{\rm s}$.90$\pm$0.00001, +7\arcdeg34\arcmin09.$''$49$\pm$0.00021) and 6.142$\pm$0.047 \mjybeam, respectively.
The major and minor axes are 0.$''$055$\pm$0.$''$061 and 0.$''$043$\pm$0.$''$050, respectively (P.A. 177\degr$\pm$2\degr).
The intensity is consistent with the previous report \citep[4.8 \mjybeam\ at 230 GHz with the 0.$''$03 beam;][]{Bjerkeli et al.(2019)}, if the slight difference of the observing frequency is considered.
As shown in the schematic picture of Figure \ref{cont_B335}(c), the disk/envelope system of this source is close to edge-on with respect to the line of sight \citep{Hirano et al.(1988), Stutz et al.(2008)}.

\subsection{Principal Component Analysis for the Cube Data (PCA-3D)}\label{sec-data}
\par
\begin{figure}[h!]
\rotatebox{90}{
\begin{minipage}{\textheight}
\centering

\includegraphics[scale=0.4]{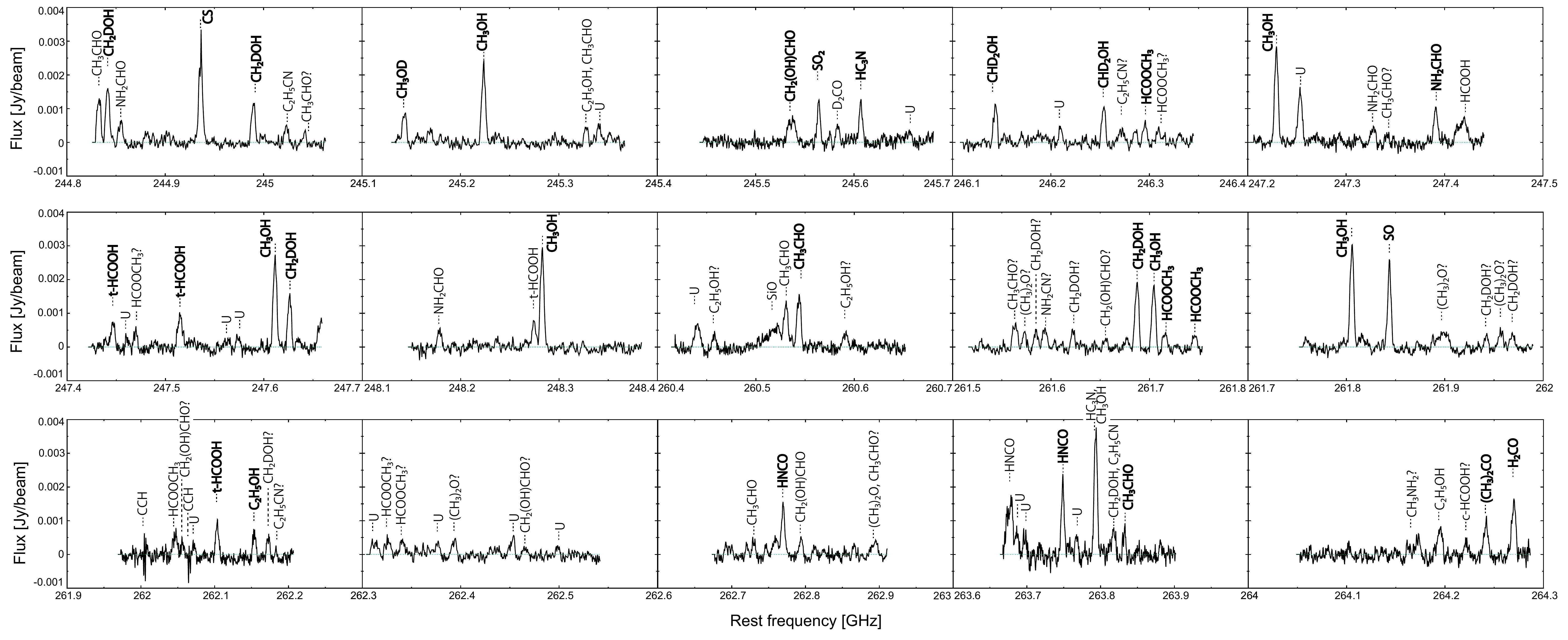}

\caption{Overview of the observed spectra for a circular region with a diameter of 0\farcs3 centered at the continuum peak position.
The lines that have not clearly identified are represented by a question mark. 'U' means an unidentified line.
\label{sp_0.3arc_B335}}
\end{minipage}}
\end{figure}

 As reported by \cite{Imai et al.(2019)}, COMs are rich in B335 and are concentrated in a vicinity of the protostar. 
In our higher-resolution (0\farcs03) observations, more than thirty lines of COMs have also been observed, as shown in the observed spectrum toward the continuum peak position (Figure \ref{sp_0.3arc_B335}). 
To our best knowledges, this is the first study of COMs other than \meta\ around a low-mass protostar at the high resolution of a few au, where the \meta\ line was observed at a similar resolution \citep{Bjerkeli et al.(2019)}.
The detected lines include oxygen-bearing molecules (e.g., \meta, CH$_3$OCHO, CH$_3$CHO, HCOOH), nitrogen-bearing molecules (HNCO, NH$_2$CHO, HC$_3$N), and deuterated species of \meta\ (CH$_2$DOH, CHD$_2$OH, and CH$_3$OD).
Here, the line identification is based on the spectral line databases, CDMS \citep{Endres et al.(2016)} and JPL catalog \citep{Pickett et al.(1998)}, except for CHD$_2$OH \citep{Mukhopadhyay(2016)} and CH$_3$OD \citep{Duan et al.(2003)}.
\par Principal Component Analysis (PCA) is a kind of multivariate analyses to find orthogonal principal coordinates representing the data and is often used to reduce the dimension of the data by transforming the original coordinates into a smaller set of orthogonal coordinates containing most of the information \citep{Jolliffe(1986)}.
It is a useful method to understand molecular distributions systematically, as reported previously \citep[e.g.,][]{Spezzano et al.(2017), Nagy et al.(2019), Okoda et al.(2020), Okoda et al.(2021)}.
In practice, we look for the orthogonal coordinates by diagonalizing the correlation matrix or the covariance matrix of the data.
The coordinates thus found are called 'principal components (PCs)', and the main features of the original data can be represented by a few PCs.
In particular, \cite{Okoda et al.(2021)} successfully perform a PCA for the molecular distributions including the velocity structures (PCA for the cube data: PCA-3D) in L483.
\par In this paper, we conduct PCA-3D of B335 to systematically reveal the small-scale chemical differentiation among these molecules in the disk/envelope system.
We select 32 lines for the PCA, as listed in Table \ref{observation_pcaB335}, and they are boldfaced in Figure \ref{sp_0.3arc_B335}.
Here, well-isolated lines without contamination are used.
The molecular lines blended with other lines are excluded (e.g., SiO and CH$_3$CHO around 260 GHz).
As well, the molecular lines which are too weak to give enough data points above the three times the rms noise level in the cube data (Table \ref{observation_pcaB335}) are excluded, because the correlation coefficients with the distributions of other lines are not well derived. 
In Table \ref{observation_pcaB335}, the molecular lines are listed in a decreasing order of the first principal component (PC1) derived later in the PCA-3D for easy comparison with the PCA results.
Figure \ref{moment_mol_B335} shows the integrated intensity (moment 0) maps of these lines.
Although all molecular lines show a round distribution around the protostar, some features are seen.
A compact intensity depression toward the protostar is probably due to the high dust opacity, as described later. 
The maps of the SO and \meta\ lines show an intensity peak at the eastern and western parts, respectively, although these distributions have been reported to be similar in some protostellar sources, L1527 \citep{Sakai et al.(2014b)} and IRAS 15398-3359 \citep{Okoda et al.(2020)} as well as in some prestellar sources, L1554 \citep{Spezzano et al.(2017)} and L1521E \citep{Nagy et al.(2019)}.
\par The data for the PCA are prepared as follows.
Since this source is known to have a small disk/envelope system, we focus on the distributions in the 0.$''$5$\times$0.$''$5 area around the protostar.
The velocity range used for the data is from -0.2 \kms\ to 14.5 \kms\ \citep[$V_{\rm sys}\sim$8.34 \kms;][]{Yen et al.(2015b)} to cover the spectral feature in the above area, where the velocity channel width is 0.7 \kms.
The velocity range is the same as that used for the moment 0 maps of Figure \ref{moment_mol_B335}.
For a fair comparison among the lines, we set the uniform beam size to be 0.$''$034$\times$0.$''$034.
These spatial area and velocity range correspond to 100$\times$100 pixels and 22 velocity channels, respectively.
\par The eigenvalues and eigenvectors of the first 7 PCs are given in Table \ref{PCA3D_B335}, where the molecular lines are ordered by PC1. 
PC1 shows the largest contribution ratio of 31.7\%.
The contribution ratios of the second and third principal components (PC2 and PC3) are 13.0 \% and 7.9 \%, respectively, and the fourth principal component (PC4) has a lower ratio of 5.0 \%.
We here discuss PC1 and PC2, where the sum of the contribution ratios is 47.7 \%.
The scree plot is shown in Figure \ref{scree_plot_B335}, where the contribution ratios of the principal components to the original data are plotted. 
The elbow in the plot suggests how many PCs mainly contribute to the original data.
Here, the molecular-line data can almost be reproduced by the first four PCs.
PC3 and PC4 also extract some characteristic features of the molecular lines.
Hence, they are briefly discussed in the Appendix \ref{pc3pc4}.

\subsection{Characteristic Features of the Principal Components}
\par The moment 0 and channel maps of the first two PCs are shown in Figure \ref{chan_B335}.
The moment 0 map of PC1 shows a round distribution around the protostar over $\sim$30 au (Figure \ref{chan_B335}(a)).
It slightly extends toward the southwest direction with an intensity depression at the continuum peak position.
In its velocity channel maps, the distribution of PC1 is clearly seen in the panels from 5.4 \kms\ to 11 \kms\  (Figure \ref{chan_B335}(c)).
The blueshifted and redshifted components can be seen mainly in the southeast and the northwest respectively, where the systemic velocity is 8.34 \kms\ \citep{Yen et al.(2015b)}.
This slight gradient can be interpreted as a rotating motion, as pointed out by \cite{Imai et al.(2019)} and \cite{Bjerkeli et al.(2019)}.
Figures \ref{pvpc1}(a) and (b) show a position-velocity (PV) diagram of PC1 along the envelope (P.A. 0$^\circ$) and outflow (P.A. 90$^\circ$) directions, respectively.
For the envelope direction, the diamond shape is seen in the velocity structure (Figure \ref{pvpc1}(a)), which is characteristic of the infalling-rotating motion, as reported previously \citep[e.g.,][]{Ohashi et al.(1997), Sakai et al.(2014b), Oya et al.(2016)}.
This result is consistent with the report by \cite{Imai et al.(2019)}.
On the other hand, in the 9.6 \kms\ and 11.0 \kms\ panels of Figure \ref{chan_B335}(c), we can see the intensity peak at the northwestern side of the continuum peak.
Since the eastern side of the disk/envelope system faces to us, the enhancement of the northwestern part is not explained by the inclination effect obscuring the emission from the backside: the eastern part of the disk/envelope structure should be more bright, if its inclination angle is considered.
According to \cite{Bjerkeli et al.(2019)}, the SiO emission traces the foot of the outflow and has an intensity peak at the western part of about 0\farcs03 from the continuum peak.
It is reported to be shocked region caused by the outflow impact \citep{Bjerkeli et al.(2019)}.
Hence, the velocity structure of PC1 would also be affected by the outflow motion.
This sign can also be confirmed in the PV diagram along the outflow direction (Figure \ref{pvpc1}(b)).
The velocity seems to be higher as an increasing distance from the continuum peak on the western side.
Figure \ref{pcspec_B335} shows the spectral line profiles of PC1 and PC2, which are prepared for a circular region with a diameter of 0.$''$1 centered at the continuum peak.
The spectrum of PC1 reveals a single peak at the slightly blueshifted velocity (Figure \ref{pcspec_B335}(a)).
\par  As shown in Figure \ref{chan_B335}(b), the moment 0 map of PC2 shows a positive compact distribution in the vicinity of the protostar as well as the negative crescent distribution around it.
Note that the PCA process yields such a negative distribution in some PCs, because we look for the orthogonal axes for molecular line distributions in the process.
Its distribution is also seen clearly in the velocity channel map of PC2  (Figure \ref{chan_B335}(d)).
The spectral line profile of PC2 has a positive double-peak feature with a dip of $\sim$6 \kms (Figure \ref{pcspec_B335}(b)), where the positive compact distribution has a broader velocity width than in PC1. 
PC2 reproduces high-velocity components of the molecular lines in combination with PC1.
The molecular lines having such a feature are described in the next subsection.

\subsection{Characteristics of the Molecular Lines Extracted by the PCA}\label{pca_results}
\label{B335_pcaresult}
\par Figures \ref{load_B335}(a) and (b) show the correlation coefficients between the principal components and the cube data for each molecular line, by using
the method described previously \citep{Okoda et al.(2020), Okoda et al.(2021)}.
The numbers for the molecular lines are assigned as a decreasing order of the PC1 value.
In Figure \ref{load_B335}(a), 19 molecular lines have a correlation coefficient larger than 0.5 for PC1, where the high excitation \meta\ lines and most of the \metd\ lines show the largest value.
This suggests that most molecular lines have a similar distribution to PC1, that is roundly extended distribution around the protostar with an intensity depression at the continuum peak.
Based on the velocity structure of PC1 (Figures \ref{chan_B335}(c) and \ref{pvpc1}(a)), these lines reveal a rotating structure of the disk/envelope system.
\par 
Figure \ref{load_B335}(b) depicts the correlation between PC2 and each molecular line.
While the \meta\ lines showing the large correlation for PC1 (\#1, \#2, \#3, \#5, \#6, and \#10) are not well correlated with PC2, the nitrogen-bearing organic species, \nhhcho\ (\#13), HNCO\ (\#11 and \#18), and \hcccn\ (\#26), have a positive correlation for PC2.
This indicates different distributions between the oxygen-bearing and nitrogen-bearing species.
Furthermore, the HCOOH lines (\#19, \#21, and \#24) have a large positive correlation for PC2 and show a similar trend to the nitrogen-bearing molecular lines, despite an oxygen-bearing species.
A positive large correlation of PC2 means the compact distribution with the high-velocity component of these molecular lines.
As well, the SO$_2$ line (\#20) is also correlated with PC2 positively, indicating that its distribution is more compact than those of the oxygen-bearing organic species (Figure \ref{moment_mol_B335}(t)).
In contrast, the deuterated species of \meta\ (\metd: \#7, \#8, \#9, and \#15, CHD$_2$OH: \#17 and \#22, and CH$_3$OD: \#25), CH$_3$OCHO (\#27), and CH$_3$CHO (\#4 and \#23) lines are correlated with PC2 negatively.
This result indicates that they have a more extended distribution around the protostar.
In particular, it is remarkable that the moment 0 map of CH$_3$OD (\#25) resembles the negative crescent of PC2 (Figure \ref{moment_mol_B335}(y)).
In short, the molecular lines with a negative correlation for PC2 have an extended distribution, resulting in their narrow line width.
\par An unbiased classification of the molecular lines can be done in the PC1-PC2 biplot of Figure \ref{load_B335}(c), which shows the contributions of the principal components for each molecular-line distribution.
The cyan, blue, and light blue plots represent, \meta, deuterated \meta\ (\metd, CHD$_2$OH, and CH$_3$OD), and the other oxygen-bearing organic molecules, respectively.
The yellow plots indicate the nitrogen-bearing molecules.
The PC2 contributions of the nitrogen-bearing molecules tend to be larger than those of the 
oxygen-bearing ones except for HCOOH (\#19, \#21, and \#24), indicating the compact distribution of the nitrogen-bearing ones.
Thus, the positive and negative PC2 contributions clearly reveal the systemic differentiation between the nitrogen-bearing and oxygen-bearing molecules.
\par 
Furthermore, a small systematic difference between \meta\ and its deuterated species can be seen in the plot (Figure \ref{load_B335}(c)).
The deuterated species tend to have a larger negative contribution of PC2, indicating a more extended distribution with a narrower linewidth than \meta.
Although their extended features can be recognized by eye in the moment 0 maps (Figure \ref{moment_mol_B335}), we can show it by the PCA-3D definitively in an unbiased way. 
\par We reveal the molecular-line distributions systematically by using PCA.
The trend for the oxygen-bearing and nitrogen-bearing species is extracted as in the case of PCA-3D in L483 reported by \cite{Okoda et al.(2021)}, and the HCOOH lines are found to show a similar trend to the nitrogen-bearing molecules, \nhhcho\, HNCO, and \hcccn.

\section{Temperature Distribution}\label{Temperature Distribution}
\par In Section \ref{Chemical Differentiation}, PCA-3D reveals that the distributions of HCOOH and \nhhcho\ are more compact than those of \meta\ and \metd.
As the first step to explore the chemical processes, it is essential to derive the gas temperature in the disk/envelope system.
We select the four species, HCOOH, \nhhcho, \metd, and \meta\ for this purpose because multiple lines are detected for each of them.
The molecular lines for the derivation are listed in Table \ref{observations_b335_shock}.
In this analysis, we add some lines of HCOOH and \nhhcho, which are not used in PCA due to their weak intensities and proximity to adjacent lines.
We derive the rotation temperature at the positions along the envelope direction in Figure \ref{cont_B335}(b).

\subsection{Derivation of Temperature}\label{LTE}
\par The rotation temperatures at the 9 positions (position 1-8 of Figure \ref{cont_B335}(b) and the continuum peak position) along the envelope direction are derived under the assumption of LTE (local thermodynamic equilibrium) from the observed intensities and the velocity widths of the HCOOH, \nhhcho, \metd, and \meta\ lines (Table \ref{observations_b335_shock}).
Since the dust emission is bright particularly toward the protostar position, we explicitly consider the effect of the optical depth of the dust emission ($\tau_{\rm dust}$) as well as that of the line emission ($\tau_{\rm line}$).
For simplicity, we assume the condition that gas and dust are well mixed and the gas temperature is equal to the dust temperature.
In this case, the observed intensity ($T_{\rm obs}$) is represented as follows based on radiative transfer:

\begin{equation}
T_{\rm obs}=\frac{c^2}{2\nu^2k_{\rm B}}\ \biggl[B_{\nu}(T)+\ {\rm exp\ }\biggl\{-(\tau_{\rm line}+\tau_{\rm dust})\biggr\}\ \biggl\{B_{\nu}(T_{\rm cb})-B_{\nu}(T)\biggr\}-I_{\rm dust}\biggr],
\end{equation}
where $B_{\nu}(T)$ and $B_{\nu}(T_{\rm cb})$ are the Planck function for the source temperature at $T$ and the cosmic microwave background temperature $T_{\rm cb}$, respectively.
$\tau_{\rm line}$ represents the optical depth of the molecular line, which can be written under the assumption of LTE as:
\begin{equation}
\tau_{\rm line}=\frac{8\pi^3 S\mu^2}{3h\Delta v U(T)}\ \biggl\{{\rm exp\ }\biggl(\frac{h\nu}{k_{\rm B}T}\biggr)-1\biggr\}{\ \rm exp\ }\biggl(-\frac{E_u}{k_{\rm B}T}\biggr)\ N,
\end{equation}
where $S$ is the line strength, $\mu$ the dipole moment responsible for the transition, $h$ the Planck constant, $\Delta v$ the full width at half maximum, $U(T)$ the partition function of the molecule at the source temperature $T$, $\nu$ the frequency, $E_u$ the upper-state energy, and $N$ the column density.
On the other hand, the intensity of the dust continuum emission ($I_{dust}$) near the line frequency is given as:
\begin{equation}
I_{dust}=B_{\nu}(T)+{\rm exp}(-\tau_{\rm dust})[B_{\nu}(T_{cb})-B_{\nu}(T)].
\end{equation}
Then, if the source temperature ($T$) is derived from the multi-line analysis, we obtain $\tau_{\rm dust}$ from $I_{dust}$ as:
\begin{equation}
\tau_{\rm dust}=-\ln\biggl\{\frac{B_{\nu}(T)-I_{\rm dust}}{B_{\nu}(T)-B_{\nu}(T_{\rm cb})}\biggr\}.
\end{equation}
If we assume that $B_{\nu}(T)\gg B_{\nu}(T_{\rm cb})$, we have:
\begin{equation}
\tau_{\rm dust}\sim-\ln\biggl\{\frac{B_{\nu}(T)-I_{\rm dust}}{B_{\nu}(T)}\biggr\}.
\end{equation}
Based on these equations, we fit the observed intensities of each molecule at each position by using a non-linear least-squares method to determine the rotation temperature ($T$) and the molecular column density ($N$).
An example of the calculation results is shown in Appendix \ref{app_fitting}.

\subsection{Temperature Distribution along the Disk/envelope System}\label{temperature_distribution}
\par The intensity profiles of the molecular lines along the envelope direction (Figure \ref{cont_B335}(a)) are shown in Figure \ref{disk_temp}(a).
The intensity depression at the continuum peak is due to the high dust opacity.
The distributions of HCOOH and \nhhcho\ are more compact than those of \meta\ and \metd, as revealed by the PCA-3D results (Section \ref{pca_results}).
Hence, we derive the rotation temperatures within the offset of 0.$''$06 for all molecules.
Since \meta\ and \metd\ are extended outside that radius, we also calculate the temperatures for the positions up to the offset of 0\farcs13 to the north and the south.
The results are summarized in Table \ref{disktemp}, where the errors are the fitting errors in the least-squares analysis.
Examples of the fit at the positions of $\pm$0.$''$06 are shown in Table \ref{results_fitting} of Appendix \ref{app_fitting}.
The line optical depth is rather high, so that the column density has a large uncertainty.
In contrast, the rotation temperature is well constrained.
The derived $\tau_{\rm dust}$ for the four lines are also summarized in Table \ref{tau_dust} of Appendix \ref{app_fitting}.
It is derived to be 1.1-1.4 at the continuum peak at most.
\par Figure \ref{disk_temp}(b) shows the plot of the derived rotation temperatures along the envelope direction, where the colors represent the molecules: orange, green, yellow, and blue circles denote the rotation temperatures for HCOOH, \nhhcho, \metd, and \meta, respectively.
The error bar is shown for each value, and the purple line represents the intensity profiles of the 1.2 mm continuum, whose brightness temperature is 141 K at its peak position.

\par
For all of the four molecules, the derived rotation temperature is the highest at the continuum peak position, which is in the range of 193$-$215 K, and decreases as an increasing distance from the protostar.
Such a high temperature toward the continuum peak is often seen in hot corino sources \cite[e.g.,][]{Watanabe et al.(2017b), Jorgensen et al.(2018), De Simone et al.(2020), Bianchi et al.(2020), Oya(2020)}.
The temperatures are 154$-$195 K at the positions of $\pm$0.$''$03 for the four molecules.
These temperatures are roughly consistent with that expected from the protostellar luminosity of 1.6 $L_\odot$.
However, at the positions of $\pm$0.$''$06, the temperatures are remarkably different among the molecules.
While \meta\ and \metd\ show the high temperature of 150-165 K both in the south and north sides, the temperatures of HCOOH and \nhhcho\ are 75-76 and 112 K, respectively.
The latter two are significantly lower than the former two.
Namely, the temperatures of HCOOH and \nhhcho, which have a more compact distribution, decrease more steeply as an increasing distance from the protostar than those of \meta\ and \metd\ (Figure \ref{disk_temp}(c)).
\meta\ and \metd\ having an extended distribution show the high temperature even in the outer parts (Figures \ref{disk_temp}(b) and (c)).
The different temperature distributions suggest that the outer region is heated by some mechanisms other than the radiation from the central source.
\par This situation can be explained by the temperature structure schematically shown in Figure \ref{disk_temp}(c). 
Since the disk/envelope system of B335 is close to an edge-on configuration, the temperature derived in the above analysis is an effective one averaged over the line of sight. 
If the outer envelope is heated by some mechanisms rather than by the protostar radiation, the temperature derived for \meta\ and \metd\ along the line of sight shown in Figure \ref{disk_temp}(c) can effectively be higher than that for HCOOH and \nhhcho\ due to the contribution of the outer envelope having a higher temperature.


\subsection{An Implication of the Accretion Shock}
\par What is an origin for the above temperature structure?
Accretion shock can be a possible explanation for the difference of the rotation temperature at the positions of $\pm$0.$''$06.
In general, an infalling-rotating gas cannot fall beyond a certain radius, unless the angular momentum of the gas is extracted by some mechanisms. 
This radius is called as a centrifugal barrier, which corresponds to a half of the centrifugal radius \citep{Sakai et al.(2014b), Oya et al.(2016)}.
 The gas is stagnated in front of the centrifugal barrier, and an accretion shock is caused by the infalling gas there \citep{Sakai et al.(2017)}. 
If this picture is the case, the centrifugal barrier in B335 would roughly be estimated to be around the radius of $\sim$0.$''$06 ($\sim$10 au) from the continuum peak.
This is almost consistent with the previous estimate of $<$ 8 au based on the gas kinematics \citep{Imai et al.(2019)}.
Since the \meta\ and \metd\ emission are weaker outside the radius of 0$\farcs$1 despite the large recoverable scale of 0$\farcs$3 (Figure \ref{disk_temp}(c)), a larger radius of the centrifugal barrier would be ruled out.
 According to the PV diagram of PC1, the gas is infalling and rotating, and the line-of-sight gas velocity at the radius of 0.$''$06 is about 3 \kms\ with respect to the systemic velocity \citep[8.34 \kms:][]{Yen et al.(2015b)}.
 If this velocity can contribute to the shock, the temperature can be raised up to about 300 K for the adiabatic case \citep{DuleyandWilliams(1984), Yamamoto(2017), Aota et al.(2015), Miura et al.(2017)}.
 Hence, the accretion shock heating is indeed possible.
\par A similar idea is reported for the other protostellar source, IRAS 16293$-$2422 Source A, by \cite{Oya(2020)}.
They derived the rotation temperature of H$_2$CS along the disk/envelope system of IRAS 16293-2422 Source A at a resolution of $\sim$14 au, and found that the temperature profile within the radius of 50 au is flattened.
This feature is most likely due to the local steep rise around 50 au.
The derived profile cannot be explained by only the radiation heating by the central protostar (A1), and hence, they suggest the accretion shock as another heating mechanism.
\par If the inside and outside of the disk/envelope system are heated by the protostar radiation and the accreting gas, respectively, the radial temperature profile as a function of the distance from the protostar would have a local minimum between the inner and outer regions (Figure \ref{disk_temp}(c)).
This situation results in the ring-like structure of an enhanced temperature region surrounding the disk structure in the case of the face-on configuration of the disk/envelope system.
This is also indicated by the model study reported in \cite{Fateeva et al.(2011)}.
\par In our observation, the ring-like structure is not observed because the disk/envelope system is almost edge-on (Figure \ref{cont_B335}(c)). 
However, the derived temperatures depend on the molecules at $\pm$0$\farcs$06, where the compactly-distributed HCOOH and \nhhcho\ lines show the relatively lower temperatures, suggesting the temperature decrease around the boundary between the inner and outer regions (Figure \ref{disk_temp}(b)).
In contrast, the rotation temperature of \meta\ remains high there due to the contribution of the high temperature gas in the outer part along the line of sight.
This picture is further supported by the fact that the \meta\ temperatures at the positions of $\pm$0$\farcs$1 in the north and the south are higher than 100 K despite a weak dust emission \ (Figure \ref{disk_temp}(b) and Table \ref{disktemp}).
As well, it should be noted that the equally high temperature of \meta\ at $\pm$0$\farcs$06 on both north and south sides suggests that the temperature raise in the north would not be due to the possible outflow shock on the north-western part seen the \meta\ intensity enhancement.
\par Thus, our results suggest that the accretion shock heating occurs at the radius of 0$\farcs$06 or larger.
It is worth noting that the HCOOH and \nhhcho\ lines are not seen in the outer region despite the temperature at which the ice mantles would be sublimated.
Hence, HCOOH and \nhhcho\ may not simply be supplied from dust grains even in the inner envelope.
This point is discussed in the next section.

\section{HCOOH and Nitrogen-bearing Species}\label{HCOOH}
\par As mentioned in Section \ref{intro}, the different distributions of nitrogen-bearing COMs and oxygen-bearing COMs have been revealed in star forming regions. 
Furthermore, a specific distribution of HCOOH has also been pointed out. 
For instance, in the high-mass star forming region, Orion KL, the distribution of HCOOH is reported to be different from other oxygen-bearing COMs such as HCOOCH$_3$ and (CH$_3$)$_2$O \citep{Liu et al.(2002)}. 
\cite{Feng et al.(2015)} revealed the similarity of the distribution of HCOOH to that of nitrogen-bearing COMs in Orion KL. 
\cite{Tercero et al.(2018)} reported the differentiation among oxygen-bearing COMs within Orion KL, where the distribution of HCOOH is different from CH$_3$OH. 
Since such chemical differentiation is found in a complex structure of Orion KL on a scale of a few 1000 au, the different distributions between nitrogen-bearing COMs and oxygen-bearing COMs as well as the specific distribution of HCOOH have been discussed in terms of the physical and chemical conditions and/or evolution history of sub-components. 
\par On the other hand, we have revealed the chemical differentiation within the disk/envelope system on a few au scale in B335. 
Our result of PCA-3D shows the compact distribution of nitrogen-bearing molecules. 
Moreover, the HCOOH lines are found to have a similar distribution to them. 
Although such trends have recently been suggested for the other low-mass protostellar sources: L483 \citep{Oya et al.(2017), Okoda et al.(2021)}, S68N, and B1-c \citep{Nazari et al.(2021)}, we have revealed the differentiation in relation to the temperature structure in B335. 
For the high-mass star forming region G328.2551-0.5321, \cite{Csengeri et al.(2019)} reported the compact distribution of HCOOH and the nitrogen-bearing COMs, which is essentially consistent with our result.  
We note that, in contrast, the distributions of \meta\ and HCOOH are reported to be similar to each other in the prestellar core L1544 \citep{Jimenez-Serra et al.(2016)}.

\par 
HCOOH is the simplest carboxylic acid and has been observed toward high-mass and low-mass star-forming regions \citep[e.g.,][]{Woods et al.(1983), Liu et al.(2001), Liu et al.(2002), Bisschop et al.(2007a), Lefloch et al.(2017), Oya et al.(2017), Csengeri et al.(2019)}, protoplanetary disks \citep[e.g.,][]{Favre et al.(2018)}, the prestellar sources \citep[e.g.,][]{Irvine et al.(1990), Vastel et al.(2014), Jimenez-Serra et al.(2016)}, and comets of the solar system \citep{Biver et al.(2014)}.
Although the production of HCOOH has also been investigated experimentally \citep{Ioppolo et al.(2011)} and theoretically \citep{TielensandHagen(1982), GarrodandHerbst(2006), Aikawa et al.(2008), Garrod et al.(2008), Vasyunin et al.(2017)}, its formation process and chemical link to nitrogen-bearing species are puzzuling.
Indeed, the chemical network calculation tracing a prestellar/protostellar core by \cite{Aikawa et al.(2020)} indicates that nitrogen-bearing COMs and HCOOH appear in the same region (i.e., a hot core) as oxygen-bearing COMs: no differentiation is found.
It should be noted that the differentiation is not ascribed to the desorption temperature of these molecules formed on grain mantles.
As described in Section \ref{temperature_distribution}, the temperature of the outer envelope traced by \meta\ is 150-165 K, which is higher than that of the inner envelope traced by HCOOH and \nhhcho, 75-112 K.
Hence, ice mantles containing various COMs should have already been liberated from dust grains in the outer envelope.
However, the results of PCA-3D suggest that HCOOH and the nitrogen-bearing species, \nhhcho, HNCO, and HC$_3$N, do not appear in the gas phase outside the radius of $\sim$0$\farcs$06, even though these desorption temperatures are comparable to those of \meta\ \citep[e.g.,][]{Oya et al.(2019)}.
In other words, the other factors rather than the desorption temperature should be responsible for these molecules to be observed in the gas phase.
It may be the high-density condition, the protostellar radiation, or both of them.
Gas-phase production of nitrogen-bearing COMs and HCOOH should be considered seriously.
\par In this relation, we briefly discuss the abundance ratios of HCOOH and \nhhcho\ relative to \meta. 
At the offsets of $\pm$0$\farcs$03, the HCOOH/\meta\ and \nhhcho/\meta\ ratios are 0.02-0.03. 
These values are higher than those found in other sources L483, B1-c, and S68N, and IRAS 16293-2422 Source A : $10^{-3}-10^{-4}$ \citep{Jacobsen et al.(2019), Manigand et al.(2020), van Gelder et al.(2020)}, as shown in Table \ref{frac_abundance}.
Interestingly, we note that these ratios of B335 are comparable to those reported by the chemical network calculation \citep{Garrod et al.(2022)}: $10^{-2}$ for HCOOH/CH$_3$OH and (1-6)$\times10^{-3}$ for NH$_2$CHO/CH$_3$OH. 
The NH$_2$CHO/CH$_3$OH ratio in IRAS 16293-2422 Source A is also consistent with the result of the chemical model.
However, the other ratios shown in Table \ref{frac_abundance} are lower than the model results.
There would be the two reasons for the discrepancy. 
First, the CH$_3$OH abundance obtained in this study may be underestimated due to high optical depths of the observed lines. We consider the optical depth effect by using equation (1), but it may not be sufficient. 
Second, the observed abundances of HCOOH and NH$_2$CHO in the other sources may be underestimated due to the beam dilution effect, if their distribution is more compact than that of CH$_3$OH, as found in this study. 
In fact, the resolution is relatively low for B1-c and S68N.
For L483, the resolution of 50 au may not be enough to resolve the hot core (Table \ref{frac_abundance}).
In any case, we need more observations toward other protostars at a resolution of a few au scale to fully understand the chemical process of organic molecules.
Related experimental and theoretical studies are also awaited.

\section{Summary}
\par We have conducted the high-resolution observations of COMs in the low-mass Class 0 protostellar source B335 with ALMA.
Our major findings are summarized below.\\
\\
1. We perform PCA for the cube data of 32 molecular-line emission observed at a high angular resolution of 0\farcs03.
We find that the distributions of the nitrogen-bearing molecules, \nhhcho, HNCO, and HC$_3$N as well as HCOOH, are more compact than those of the oxygen-bearing species such as \metd\ and \meta.
Clear differentiations of COM distributions are thus identified on a few au scale around the protostar of B335.\\

\noindent
2. We derive the rotation temperatures of the disk/envelope system under the assumption of LTE condition by using the multi-line analysis of HCOOH, \nhhcho, \metd, and \meta.
Along the envelope direction, the temperatures at the radius of 0\farcs06 of \meta\ and \metd\ (150-165 K) are found to be higher than those of HCOOH and \nhhcho\ (75-112 K).
This result suggests that the outer envelope is heated by some mechanisms other than the protostellar radiation.
We propose the accretion shock as the mechanism to account for, based on the temperature distribution.\\

\noindent
3. The compact distributions of HCOOH and nitrogen-bearing molecules along with the temperature structure raise an important question on their formation mechanism.
These molecules are not observed in the outer region where the temperature is higher than the sublimation temperature of ice mantle ($>$100 K).
This result implies that they are not simply supplied from ice mantle in the central part.
Their production in the closest vicinity of the protostar should be considered.\\

\acknowledgments
We thank the reviewer for his/her invaluable comments and suggestions.
This paper makes use of the following ALMA data set:
ADS/JAO.ALMA\# 2018.1.01311.S (PI: Muneaki Imai). ALMA is a partnership of the ESO (representing its member states), the NSF (USA) and NINS (Japan), together with the NRC (Canada) and the NSC and ASIAA (Taiwan), in cooperation with the Republic of Chile.
The Joint ALMA Observatory is operated by the ESO, the AUI/NRAO, and the NAOJ. The authors thank to the ALMA staff for their excellent support.
This project is supported by a Grant-in-Aid from Japan Society for the Promotion of Science (KAKENHI: No. 18H05222, 19H05069, 19K14753, and 20J13783.
Y. Okoda thanks the Advanced Leading Graduate Course for Photon Science (ALPS), Japan Society for the Promotion of Science (JSPS), and RIKEN Special Postdoctoral Researcher Program (Fellowships) for financial support.

{}

\begin{table}[ht]
\caption{\centering{Observation Parameters$^a$}\label{ALMA_observations_B335}}
\scalebox{0.9}{
\begin{tabular}{lcccccc}
\hline \hline
Execution block &1$^b$&2$^c$&3$^d$&4$^e$	\\
\hline
Observation date			&  2019 June 10 &2019 June 12&2019 June 13 &2019 June 23 \\
Time on Source (minute)	& 45.05& 44.93&45.12&45.10 \\
Number of antennas			& 49&44&47&48\\
Observation frequency (GHz)& \multicolumn{4}{c}{244.8-264.0}\\
Primary beam width  ($''$)	& \multicolumn{4}{c}{23.7}						\\
Total bandwidth (GHz) 		&\multicolumn{4}{c}{0.234}\\
Spectral channel width (GHz)&\multicolumn{4}{c}{0.238}\\
Continuum bandwidth (GHz)	&\multicolumn{4}{c}{0.234}\\
Baseline range (m) 		&\multicolumn{4}{c}{83$-$16196}			\\
Bandpass calibrator			 &\multicolumn{4}{c}{J1924-2914}	\\
Phase calibrator				&\multicolumn{4}{c}{J1938+0448}	\\
Flux calibrator				&\multicolumn{4}{c}{J1924-2914}	\\
Pointing calibrator			&\multicolumn{4}{c}{J1924-2914}			\\
$\sigma$ (mJy beam$^{-1}$channel$^{-1}$)&\multicolumn{4}{c}{1.0}\\
\hline
\end{tabular}}
\begin{flushleft}
\tablecomments{
$^a$ These observations are conducted with the Band 6 receiver of ALMA. 
$^b$ uid://A002/Xdd7b18/X2f30
$^c$ uid://A002/Xdd7b18/Xa010
$^d$ uid://A002/Xdd7b18/X66c
$^e$ uid://A002/Xdd7b18/X73f4
}
\end{flushleft}
\end{table}

\begin{longrotatetable}
\begin{table}[ht]
\centering
\caption{\centering{Observed Molecular Lines for PCA$^a$} \label{observation_pcaB335}}
\scalebox{0.55}{
\begin{tabular}{cccccccccccc}
\hline \hline
 Number& Molecule&Transition & Frequency & $S \mu^2$ & $E_{\rm u}$$k^{-1}$ & Original beam size&$\sigma$ (Cube)&$\sigma$ (Moment 0)&Bandwidth&Resolution\\
  &&&(GHz)&($D^2$)&(\rm K)&&(\mjybeam )&(\mjybeam\kms)&(MHz)&(kHz)\\
\hline  
1 & CH$_3$OH & 18$_{3,15}-$18$_{2,16}$, A & 247.610918 & 69.431 & 447 & 0.$''$030$\times$0.$''$023\ (P.A. -32$^\circ$)  & 0.6 & 3&238&244 \\
2 & CH$_3$OH & 21$_{3,18}-$21$_{2,19}$, A & 245.223019 & 82.489 & 586 & 0.$''$030$\times$0.$''$023\ (P.A. -32$^\circ$)  & 0.6 & 3 &237&488 \\
3 & CH$_3$OH & 12$_{6,7}-$13$_{5,8}$, E & 261.704409 & 8.5234 & 360 & 0.$''$029$\times$0.$''$022\ (P.A. -29$^\circ$)  & 0.6 & 3 &238&489 \\
4 & CH$_3$CHO & 14$_{1,14}-$13$_{1,13}$, A & 260.5440195 & 175.9719 & 96 & 0.$''$030$\times$0.$''$023\ (P.A. -32$^\circ$)  & 0.7 & 3 &238&244 \\
5 & CH$_3$OH & 17$_{3,14}-$17$_{2,15}$, A & 248.282424 & 65.259 & 405 & 0.$''$030$\times$0.$''$023\ (P.A. -32$^\circ$)  & 0.7 & 3&238&490  \\
6 & CH$_3$OH & 4$_{2,2}-$5$_{1,5}$, A & 247.228587 & 4.344 & 61 & 0.$''$030$\times$0.$''$023\ (P.A. -32$^\circ$)  & 0.6 & 3&238&244 \\
7 & CH$_2$DOH & 4$_{2,2}-$4$_{1,3}$, e$_0$ & 244.8411349 & 2.54 & 38 & 0.$''$030$\times$0.$''$023\ (P.A. -32$^\circ$)  & 0.6 & 3 &237&488\\
8 & CH$_2$DOH & 5$_{2,4}-$5$_{1,5}$, e$_0$ & 261.6873662 & 4.006 & 48 & 0.$''$029$\times$0.$''$022\ (P.A. -29$^\circ$)  & 0.6 & 3 &238&489 \\
9 & CH$_2$DOH & 3$_{2,1}-$3$_{1,2}$, e$_0$ & 247.6257463 & 2.36 & 29 & 0.$''$030$\times$0.$''$023\ (P.A. -32$^\circ$)  & 0.6 & 3 &238&244 \\
10 & CH$_3$OH & 2$_{1,1}-$1$_{0,1}$, E & 261.805675 & 5.336 & 28 & 0.$''$029$\times$0.$''$022\ (P.A. -29$^\circ$)  & 0.6 & 3&238&489 \\
11 & HNCO & 12$_{0,12}-$11$_{0,11}$ & 263.748625 & 29.956 & 82 & 0.$''$028$\times$0.$''$022\ (P.A. -31$^\circ$)  & 0.8 & 3 &237&488 \\
12 & CS & 5$-$4 & 244.9355565 & 19.2 & 35 & 0.$''$030$\times$0.$''$023\ (P.A. -32$^\circ$)  & 0.5 & 3 &237&488\\
13 & NH$_2$CHO & 12$_{0,12}-$11$_{0,11}$ & 247.390719 & 156.32 & 78 & 0.$''$030$\times$0.$''$023\ (P.A. -32$^\circ$)  & 0.6 & 3&238&244  \\
14 & SO & $J_N=$7$_{6}-$6$_{5}$ & 261.843721 & 16.4 & 48 & 0.$''$029$\times$0.$''$022\ (P.A. -29$^\circ$)  & 0.6 & 3 &238&489\\
15 & CH$_2$DOH & 10$_{2,8}-$10$_{1,9}$, o$_1$ & 244.9888456 & 3.439 & 153 & 0.$''$030$\times$0.$''$023\ (P.A. -32$^\circ$)  & 0.6 &  3&237&488\\
16 & H$_2$CO & 10$_{1,9}-$10$_{1,10}$ & 264.27014 & 3.1665 & 210 & 0.$''$028$\times$0.$''$022\ (P.A. -31$^\circ$)  & 1.0 & 5&237&488  \\
17 & CHD$_2$OH & $J=$6$-$5, $K=$1+, o$_1$ & 246.1432950 & 4.8216& 53& 0.$''$030$\times$0.$''$023\ (P.A. -32$^\circ$)  & 0.6 & 3&237&488  \\
18 & HNCO & 12$_{1,12}-$11$_{1,11}$ & 262.7687626 & 29.413 & 125 & 0.$''$029$\times$0.$''$022\ (P.A. -31$^\circ$)  & 0.7 & 3&238&489 \\
19 & HCOOH & 11$_{5,7}-$10$_{5,6}$ and 11$_{5,6}-$10$_{5,5}$ & 247.5139713 & 17.633 & 151 & 0.$''$030$\times$0.$''$023\ (P.A. -32$^\circ$)  & 0.6 & 3&238&244  \\
20 & SO$_2$ & 10$_{3,7}-$10$_{2,8}$ & 245.5634219 & 14.513 & 73 & 0.$''$030$\times$0.$''$023\ (P.A. -32$^\circ$)  & 0.6 & 3 &238&244\\
21 & HCOOH & 12$_{0,12}-$11$_{0,11}$ & 262.103481 & 24.157 & 83 & 0.$''$029$\times$0.$''$022\ (P.A. -29$^\circ$)  & 0.6 & 3&238&244 \\
22 & CHD$_2$OH & $J=$6$-$5, $K=$1+, e$_0$ &   246.2530390 & 4.7596  & 45 & 0.$''$030$\times$0.$''$023\ (P.A. -32$^\circ$)  & 0.5 & 3&237&488  \\
23 & CH$_3$CHO & 14$_{0,14}-$13$_{0,13}$, E, $v_t=$1 & 263.8319277 & 175.5781 & 300 & 0.$''$028$\times$0.$''$022\ (P.A. -31$^\circ$)  & 0.8 & 3 &237&488 \\
24 & HCOOH & 11$_{6,6}-$10$_{6,5}$ and 11$_{6,5}-$10$_{6,4}$ & 247.4462429 & 15.614 & 186 & 0.$''$030$\times$0.$''$023\ (P.A. -32$^\circ$)  & 0.6 & 3 &238&244\\
25 & CH$_3$OD & 5$_1-$4$_0$, E & 245.142988 & - & - & 0.$''$030$\times$0.$''$023\ (P.A. -32$^\circ$)  & 0.6 & 3 &237&488 \\
26 & HC$_3$N & 27$-$26 & 245.6063199 & 375.98 & 165 & 0.$''$030$\times$0.$''$023\ (P.A. -32$^\circ$)  & 0.6 & 3 &238&244 \\
27 & CH$_3$OCHO & 21$_{7,14}-$20$_{7,13}$, A & 261.74658 & 99.483 & 170 & 0.$''$029$\times$0.$''$022\ (P.A. -29$^\circ$)  & 0.5 & 3&238&489 \\
28 & C$_2$H$_5$OH & 13$_{2,12}-$12$_{1,11}$ & 262.1542443 & 12.186 & 81 & 0.$''$029$\times$0.$''$022\ (P.A. -29$^\circ$)  & 0.7 & 3 &238&244 \\
29 & CH$_3$OCHO & 21$_{7,14}-$20$_{7,13}$, E & 261.715518 & 97.5789 & 170 & 0.$''$029$\times$0.$''$022\ (P.A. -29$^\circ$)  & 0.6 & 3 &238&489 \\
30 & CH$_3$COCH$_3$ & 14$_{11,3}-$13$_{10,4}$ & 264.2428888 & 126.0112 & 90 & 0.$''$028$\times$0.$''$022\ (P.A. -31$^\circ$)  & 0.9 & 3 &237&488 \\
31 & CH$_2$OHCHO & 7$_{7,1}-$6$_{6,0}$ and 7$_{7,0}-$6$_{6,1}$ & 245.5362413 & 35.102 & 45 & 0.$''$030$\times$0.$''$023\ (P.A. -32$^\circ$)  & 0.6 & 3&238&244  \\
32 & CH$_3$OCHO & 20$_{11,10}-$19$_{11,9}$, E & 246.308272 & 74.4142 & 204 & 0.$''$030$\times$0.$''$023\ (P.A. -32$^\circ$)  & 0.6 & 3  &237&488\\
 \hline
\hline
\end{tabular}
}
\begin{flushleft}
\tablecomments{
$^a$ Line parameters are taken from CDMS \citep{Endres et al.(2016)} and JPL \citep{Pickett et al.(1998)} except for CH$_3$OD. 
$^b$ The parameter for CH$_3$OD is taken from \cite{Duan et al.(2003)}. 
The root-mean-square noise ($\sigma$) and the beam size are based on the observation data.
The spectral bandwidth and the channel width are taken from the header of each image data.
The molecular lines are ordered by PC1 of PCA-3D in Table \ref{PCA3D_B335} for consistency in the numbering of molecular lines.}
\end{flushleft}
\end{table}
\end{longrotatetable}

\begin{table}[ht]
\caption{Eigenvectors of the Principal Components and Their Eigenvalues for PCA-3D$^a$ \label{PCA3D_B335}}
\scalebox{0.8}{
\begin{tabular}{ccccccccccccc}
\hline \hline
Number & Molecule & PC1 & PC2 & PC3 & PC4 & PC5 & PC6 & PC7 \\
\hline
1 & CH$_3$OH (18$_{3,15}-$18$_{2,16}$, A) & 0.276 & -0.007 & -0.123 & 0.064 & 0.089 & 0.012 & 0.037 \\
2 & CH$_3$OH (21$_{3,18}-$21$_{2,19}$, A) & 0.269 & 0.029 & 0.021 & 0.041 & -0.083 & 0.006 & 0.03 \\
3 & CH$_3$OH (12$_{6,7}-$13$_{5,8}$, E) & 0.261 & -0.056 & 0.064 & -0.102 & 0.003 & -0.069 & 0.034 \\
4 & CH$_3$CHO (14$_{1,14}-$13$_{1,13}$, A) & 0.24 & -0.112 & 0.092 & -0.091 & -0.09 & -0.041 & -0.126 \\
5 & CH$_3$OH (17$_{3,14}-$17$_{2,15}$, A) & 0.24 & -0.008 & -0.211 & 0.145 & 0.044 & 0.054 & -0.124 \\
6 & CH$_3$OH (4$_{2,2}-$5$_{1,5}$, A) & 0.235 & 0.008 & -0.241 & 0.038 & -0.031 & 0.117 & 0.004 \\
7 & CH$_2$DOH (4$_{2,2}-$4$_{1,3}$, e$_0$) & 0.234 & -0.176 & 0.048 & 0.02 & -0.021 & 0.013 & 0.017 \\
8 & CH$_2$DOH (5$_{2,4}-$5$_{1,5}$, e$_0$) & 0.23 & -0.161 & -0.043 & 0.005 & 0.09 & -0.083 & -0.096 \\
9 & CH$_2$DOH (3$_{2,1}-$3$_{1,2}$, e$_0$) & 0.224 & -0.17 & 0.037 & -0.021 & 0.129 & -0.091 & -0.178 \\
10 & CH$_3$OH (2$_{1,1}-$1$_{0,1}$, E) & 0.208 & -0.027 & -0.314 & -0.014 & -0.191 & 0.03 & -0.184 \\
11 & HNCO (12$_{0,12}-$11$_{0,11}$) & 0.202 & 0.263 & 0.012 & -0.077 & -0.042 & -0.063 & 0.135 \\
12 & CS (5$-$4) & 0.192 & -0.032 & -0.27 & 0.02 & 0.003 & -0.041 & 0.137 \\
13 & NH$_2$CHO (12$_{0,12}-$11$_{0,11}$) & 0.188 & 0.322 & 0.136 & -0.058 & 0.053 & 0.034 & 0.009 \\
14 & SO ($J_N=$7$_{6}-$6$_{5}$) & 0.176 & 0.041 & -0.247 & 0.136 & -0.061 & -0.229 & 0.186 \\
15 & CH$_2$DOH (10$_{2,8}-$10$_{1,9}$, o$_1$) & 0.159 & -0.184 & 0.214 & -0.142 & -0.116 & 0.156 & 0.067 \\
16 & H$_2$CO (10$_{1,9}-$10$_{1,10}$) & 0.159 & 0.056 & 0.207 & -0.138 & 0.067 & 0.111 & 0.006 \\
17 & CHD$_2$OH ($J=$6$-$5, $K=$1+, o$_1$) & 0.159 & -0.207 & 0.082 & -0.089 & 0.118 & 0.258 & 0.042 \\
18 & HNCO (12$_{1,12}-$11$_{1,11}$) & 0.157 & 0.176 & -0.168 & 0.038 & -0.071 & 0.187 & 0.065 \\
19 & HCOOH (11$_{5,7}-$10$_{5,6}$ and 11$_{5,6}-$10$_{5,5}$) & 0.156 & 0.279 & 0.117 & 0.008 & 0.128 & 0.242 & -0.027 \\
20 & SO$_2$ (10$_{3,7}-$10$_{2,8}$) & 0.155 & 0.211 & 0.077 & 0.144 & -0.134 & -0.38 & 0.126 \\
21 & HCOOH (12$_{0,12}-$11$_{0,11}$) & 0.145 & 0.282 & 0.23 & -0.2 & 0.082 & 0.027 & -0.029 \\
22 & CHD$_2$OH ($J=$6$-$5, $K=$1+, e$_0$) & 0.134 & -0.251 & 0.167 & 0.092 & 0.036 & 0.031 & 0.064 \\
23 & CH$_3$CHO (14$_{0,14}-$13$_{0,13}$, E, $v_t=$1) & 0.122 & -0.179 & -0.132 & -0.293 & -0.139 & -0.146 & -0.366 \\
24 & HCOOH (11$_{6,6}-$10$_{6,5}$ and 11$_{6,5}-$10$_{6,4}$) & 0.119 & 0.287 & 0.283 & -0.132 & 0.105 & 0.012 & -0.015 \\
25 & CH$_3$OD (5$_1-$4$_0$, E) & 0.118 & -0.278 & 0.165 & 0.278 & 0.019 & 0.042 & 0.129 \\
26 & HC$_3$N (27$-$26) & 0.114 & 0.268 & -0.07 & 0.278 & -0.055 & -0.071 & 0.142 \\
27 & CH$_3$OCHO (21$_{7,14}-$20$_{7,13}$, A) & 0.114 & -0.265 & 0.217 & 0.032 & 0.236 & 0.025 & 0.451 \\
28 & C$_2$H$_5$OH (13$_{2,12}-$12$_{1,11}$) & 0.089 & 0.048 & 0.274 & 0.381 & -0.245 & -0.044 & -0.319 \\
29 & CH$_3$OCHO (21$_{7,14}-$20$_{7,13}$, E) & 0.048 & -0.028 & 0.094 & -0.537 & -0.42 & -0.123 & 0.043 \\
30 & CH$_3$COCH$_3$ (14$_{11,3}-$13$_{10,4}$) & 0.034 & 0.008 & 0.17 & 0.136 & 0.395 & -0.234 & -0.52 \\
31 & CH$_2$OHCHO (7$_{7,1}-$6$_{6,0}$ and 7$_{7,0}-$6$_{6,1}$) & 0.016 & 0.093 & -0.232 & -0.095 & 0.215 & 0.557 & -0.156 \\
32 & CH$_3$OCHO (20$_{11,10}-$19$_{11,9}$, E) & -0.019 & -0.022 & 0.213 & 0.294 & -0.545 & 0.384 & -0.115 \\
\hline
\hline
&Eigenvalues&10.27 & 4.212 & 2.549&  1.618&  1.482  &1.242&  1.165  \\
&Contribution ratio (\%) &31.7& 13.0  & 7.9  &5.0   &4.6  &3.8 & 3.6 \\
\hline

\end{tabular}
}
\begin{flushleft}
\tablecomments{
\item$^a$ These values are also called as 'loadings'.}
\end{flushleft}
\end{table}

\begin{longrotatetable}
\begin{table}[h!]
\caption{\centering{Molecular Lines for the Derivation of the Temperarture$^a$}\label{observations_b335_shock}}
\scalebox{0.9}{
\begin{tabular}{ccccccccc}
\hline\hline
Molecule&Transition & Frequency & $S \mu^2$ & $E_{\rm u}$$k^{-1}$ & Beam size&$\sigma$ \\
 &&(GHz)&$(D^2)$&(K)&&(\mjybeam$\cdot$ \kms)\\
 \hline
 HCOOH 
		&11$_{6,6}$$-$10$_{6,5}$ &247.4462429$^{b}$&15.614&186&0.$''$030$\times$0$.''$ 023 (P.A. -32$^{\circ}$)&3\\
		&11$_{6,5}$$-$10$_{6,4}$&247.4462439$^{b}$&15.614&186&\\
		&11$_{5,7}$$-$10$_{5,6}$ &247.5139713$^{b}$ & 17.633&151&0.$''$030$\times$0$.''$ 023 (P.A. -32$^{\circ}$)&3\\
		&11$_{5,6}$$-$10$_{5,5}$& 247.5141176$^{b}$  &17.633&151\\
		&11$_{3,8}$$-$10$_{3,7}$&248.2744893&20.573&100&0.$''$030$\times$0$.''$ 023 (P.A. -31$^{\circ}$)&2\\
		&12$_{0,12}$$-$11$_{0,11}$&262.1034810&24.157&83&0.$''$029$\times$0.$''$022 (P.A. -29$^{\circ}$)&3\\
\nhhcho&13$_{0,13}$$-$12$_{1,12}$&244.8542130& 6.6085&91&0.$''$030$\times$0$.''$ 023 (P.A. -32$^{\circ}$)&2\\
		&12$_{0,12}$$-$11$_{0,11}$, $v_{12}=$1&247.3273220 & 156.34&494&0.$''$030$\times$0$.''$ 023 (P.A. -32$^{\circ}$)&2\\
		&12$_{0,12}$$-$11$_{0,11}$&247.3907190 &156.32&78&0.$''$030$\times$0$.''$ 023 (P.A. -32$^{\circ}$)&2\\
\metd&4$_{2,2}$$-$4$_{1,3}$, e$_0$&244.8411349 &2.540& 38&0.$''$030$\times$0$.''$ 023 (P.A. -32$^{\circ}$)&2\\ 
		&10$_{2,8}$$-$10$_{1,9}$, o$_1$&244.9888456&3.439& 153&0.$''$030$\times$0$.''$ 023 (P.A. -32$^{\circ}$)&2\\
		&3$_{2,1}$$-$3$_{1,2}$, e$_0$ &247.6257463&2.360& 29&0.$''$030$\times$0$.''$ 023 (P.A. -32$^{\circ}$)&2\\
		&5$_{2,4}$$-$5$_{1,5}$, e$_0$&261.6873662&4.006& 48&0.$''$029$\times$0.$''$022 (P.A. -29$^{\circ}$)&3\\
\meta&21$_{3,18}$$-$21$_{2,19}$, A& 245.2230190&82.489 &586&0.$''$030$\times$0$.''$ 023 (P.A. -32$^{\circ}$)&3\\
		&4$_{2,2}$$-$5$_{1,5}$, A & 247.2285870 &4.3444& 61&0.$''$030$\times$0$.''$ 023 (P.A. -32$^{\circ}$)&3\\
		&18$_{3,15}$$-$18$_{2,16}$, A & 247.6109180&  69.431&447&0.$''$030$\times$0.$''$ 023 (P.A. -32$^{\circ}$)&3\\
		&17$_{3,14}$$-$17$_{2,15}$, A & 248.2824240& 65.259& 405&0.$''$030$\times$0.$''$ 023 (P.A. -31$^{\circ}$)&3 \\
		&12$_{6,7}$$-$13$_{5,8}$, E & 261.7044090& 8.5234&360&0.$''$029$\times$0.$''$022 (P.A. -29$^{\circ}$)&3\\
		&2$_{1,1}$$-$1$_{0,1}$, E & 261.8056750 & 5.336&28&0.$''$029$\times$0.$''$022 (P.A. -29$^{\circ}$)&3\\


\hline
\end{tabular}
}
\begin{flushleft}
\tablecomments{
$^a$ Line parameters are taken from CDMS \citep{Endres et al.(2016)} and JPL \citep{Pickett et al.(1998)}. 
The rms noise ($\sigma$) and the beam size are based on the observation data.
$^b$ Blended.}
\end{flushleft}
\end{table}
\end{longrotatetable}

\begin{table}[h!]
\centering
\caption{Rotation Temperature and Column Density along the Envelope Direction$^a$\label{disktemp}}
\scalebox{0.7}{
\begin{tabular}{cccccccccc}
\hline\hline
\multicolumn{10}{c}{Temperature [K]}  \\
Position & Offset ($''$) & \multicolumn{2}{c}{HCOOH}  & \multicolumn{2}{c}{NH$_2$CHO} &   \multicolumn{2}{c}{\metd} &  \multicolumn{2}{c}{\meta}  \\
\hline
1 & 0.13 & - & - & - & - & 44 & (4) & 102 & (0.2) \\
2 & 0.10 & - & - & - & - & 74 & (3) & 110 & (5) \\
3 & 0.06 & 75 & (7) & 112 & (15) & 150 & (21) & 156 & (7) \\
4 & 0.03 & 187 & (19) & 195 & (5) & 161 & (3) & 194 & (9) \\
Continuum peak & 0 & 213 & (5) & 215 & (8) & 193 & (11) & 211 & (3) \\
5 & -0.03 & 180 & (13) & 159 & (8) & 154 & (6) & 176 & (5) \\
6 & -0.06 & 76 & (3) & 112 & (5) & 165 & (27) & 160 & (5) \\
7 & -0.10 & - & - & - & - & 127 & (13) & 136 & (9) \\
8 & -0.13& - & - & - & - & 65 & (5) & 110 & (13) \\
\hline
 \multicolumn{10}{c}{Column density [10$^{18}$ cm$^{-2}$]}  \\
Position & Offset ($''$) & \multicolumn{2}{c}{HCOOH}   & \multicolumn{2}{c}{NH$_2$CHO} &   \multicolumn{2}{c}{CH$_2$DOH} &  \multicolumn{2}{c}{CH$_3$OH}   \\
\hline
1 & 0.13 & - & - & - & - & 1.2 & (0.9) & 1.6 & (0.01) \\
2 & 0.10 & - & - & - & - & 0.67 & (0.15) & 4.6 & (1.0) \\
3 & 0.06 & 0.17 & (0.09) & 0.17 & (0.07) & 0.89 & (0.07) & 7.9 & (2.5) \\
4 & 0.03 & 0.54 & (0.21) & 0.31 & (0.03) & 2.5 & (0.1) & 17 & (9) \\
Continuum peak & 0 & 1.1 & (0.4) & 0.18 & (0.05) & 2.5 & (1.4) & 8.5 & (2.3) \\
5 & -0.03 & 0.38 & (0.07) & 0.37 & (0.09) & 2.2 & (0.3) & 16 & (7) \\
6 & -0.06 & 0.22 & (0.04) & 0.17 & (0.02) & 1.1 & (0.1) & 8.6 & (1.9) \\
7 & -0.10 & - & - & - & - & 0.83 & (0.15) & 4.8 & (1.6) \\
8 & -0.13 & - & - & - & - & 0.49 & (0.35) & 2.4 & (1.8) \\
\hline
\end{tabular}
}
\begin{flushleft}
\tablecomments{
$^a$ The rotation temperatures and the column densities derived from a multi-line analysis (Section \ref{LTE}).
Numbers in parentheses denote the fitting error in the least-square analysis.
The envelope direction and the position numbers are shown in Figure \ref{cont_B335}(b).}
\end{flushleft}
\end{table}

\begin{table}[h!]
\centering
\caption{Fractional Abundances Relative to \meta\ around the Protostars\label{frac_abundance}}
\scalebox{0.8}{
\begin{tabular}{cccccccc}
\hline\hline
Source & B335$^a$ & L483$^b$ & B1-c$^c$ & S68N$^c$ & IRAS 16293 A$^d$ \\
\hline
$\lbrack$HCOOH$\rbrack$/$\lbrack$\meta$\rbrack$ & 0.02-0.03 & - & 4.0$\times$10$^{-4}$ & 7.0$\times$10$^{-4}$ & 1.0$\times$10$^{-3}$  \\
$\lbrack$\nhhcho$\rbrack$/$\lbrack$\meta$\rbrack$ & 0.02 & 5.9$\times$10$^{-4}$ & - & - & 1.5$\times$10$^{-3}$ \\
\hline
Resolution (au)&5 & 50&144  &196   &70  \\
\hline
\end{tabular}
}
\begin{flushleft}
\tablecomments{
\item$^a$ This work. These values are derived from the column densities at the offsets of $\pm$0\farcs03, because the optical depth is high at the continuum peak.
$^b$ \cite{Jacobsen et al.(2019)}. 
$^c$ B1-c in the Perseus Barnard 1 cloud and Serpens S68N \citep{van Gelder et al.(2020)}.
$^d$ IRAS 16293-2422 Souce A \citep{Manigand et al.(2020)}.
}
\end{flushleft}
\end{table}

\begin{figure}[h!]
\centering

\includegraphics[scale=0.23]{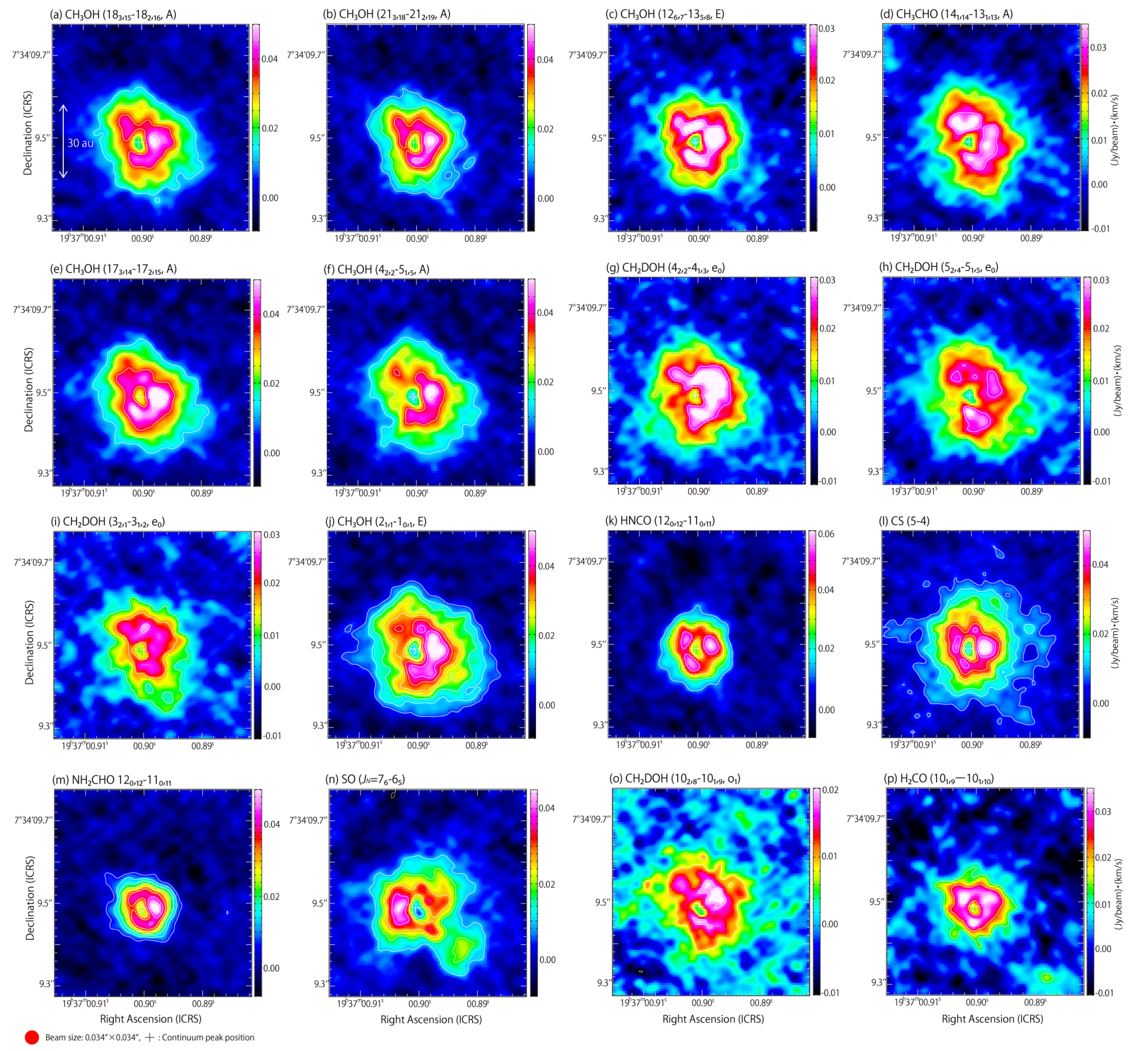}

\caption{Moment 0 maps of the 32 molecular lines, which are used in PCA-3D. 
A uniform spatial resolution of 0.$''$034$\times$0.$''$034 is applied for these images. 
The cross marks show the continuum peak position: ($\alpha_{2000}$, $\delta_{2000}$) = (19$^{\rm h}$37$^{\rm m}$00$^{\rm s}$.90$\pm$0.00001, +7\arcdeg34\arcmin09.$''$49$\pm$0.00021).
 The order of (a)-(af) is the same as that in Table \ref{observation_pcaB335} (1-32).
The integrated velocity range is from -0.2 \kms\ to 14.5 \kms. 
Contour levels are every 3$\sigma$ from 3$\sigma$, where $\sigma$ is listed in Table \ref{observation_pcaB335}. \label{moment_mol_B335}}
\end{figure}
\addtocounter{figure}{-1}
\begin{figure}[h!]
\centering

\includegraphics[scale=0.23]{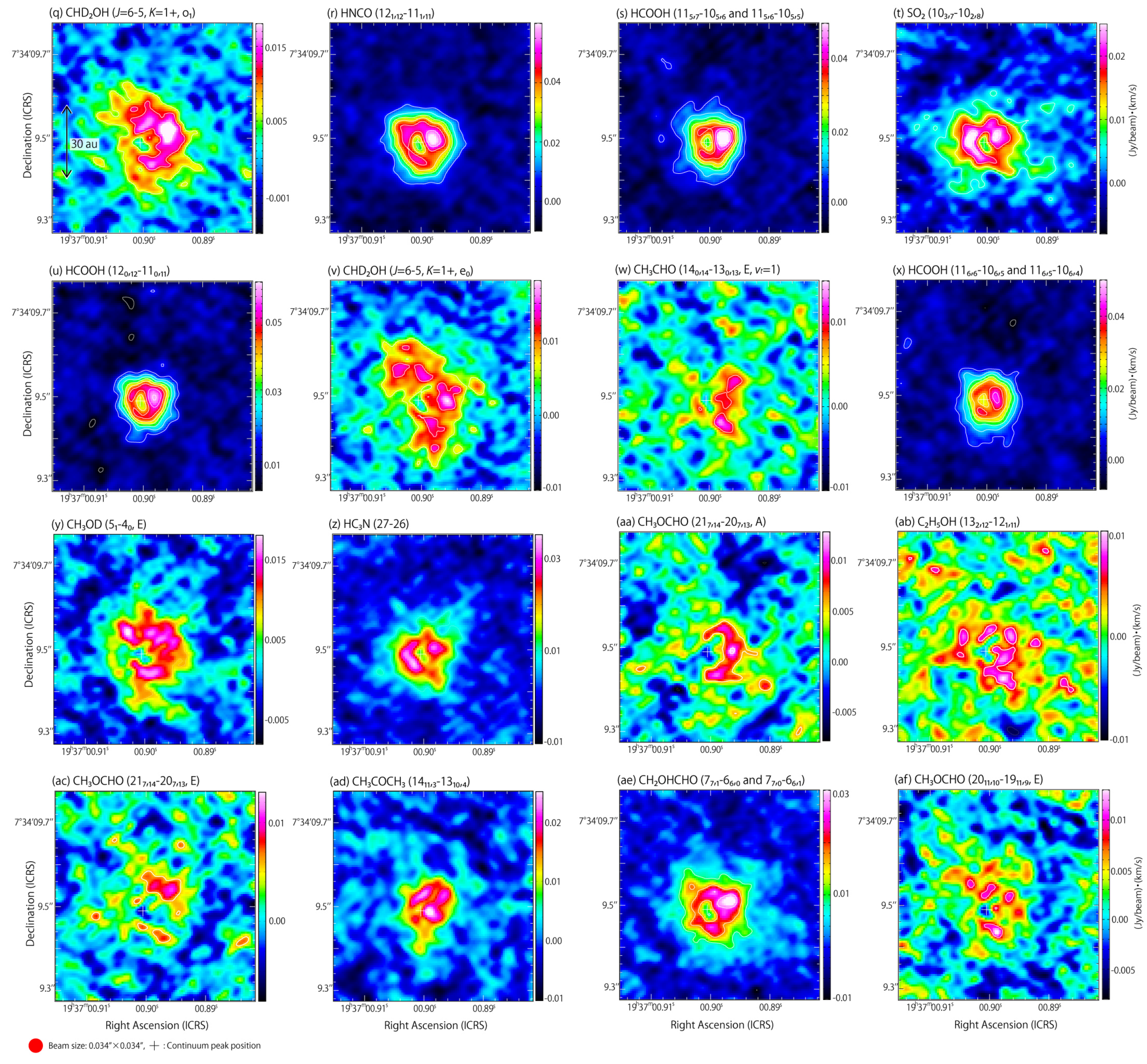}

\caption{(Continued.)\label{moment_mol_B335}}
\end{figure}

\begin{figure}[h!]
\centering

\includegraphics[scale=0.7]{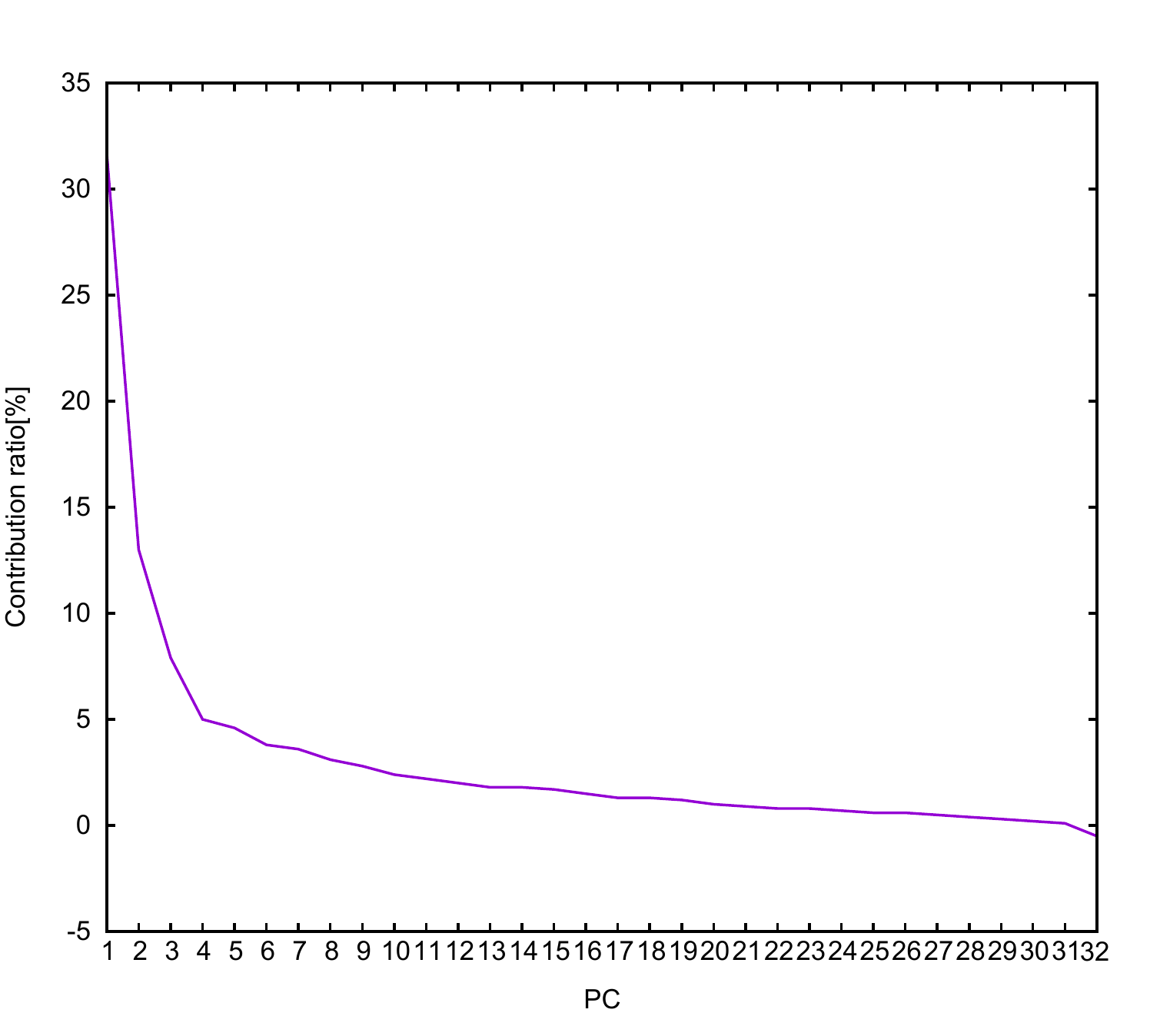}

\caption{Scree plots of the contribution ratios of the principal components. The elbow seen at PC4 in the plot suggests that the first four PCs mainly contribute to the original data.
In other words, the molecular-line data can almost be reproduced by only the first four PCs.\label{scree_plot_B335}}
\end{figure}

\begin{figure}[h!]
\rotatebox{90}{
\begin{minipage}{\textheight}
\centering

\includegraphics[scale=0.4]{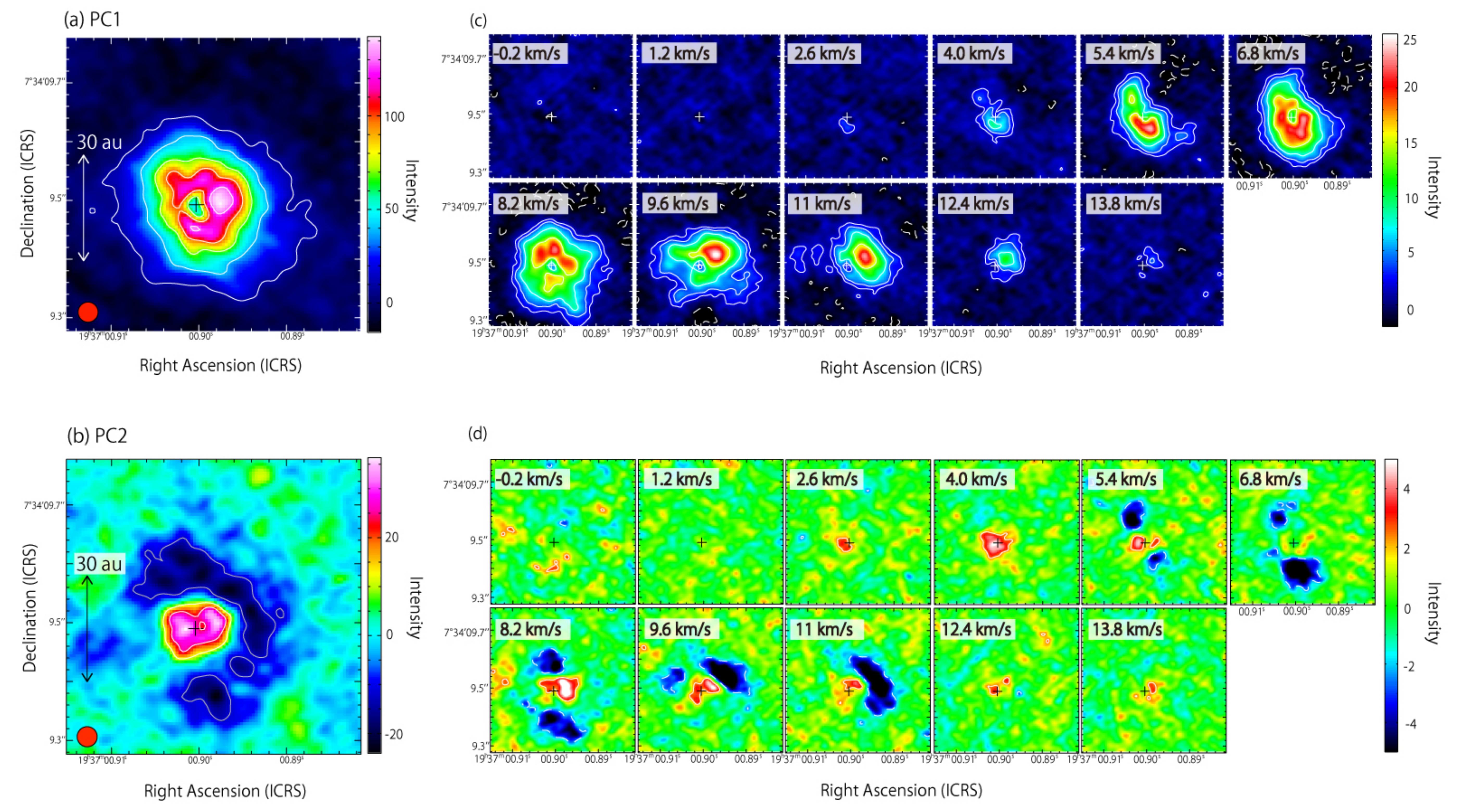}
\caption{(a, b) Moment 0 maps of the first two principal components in PCA-3D. 
The velocity range of integration is from -0.2 \kms\ to 14.5 \kms.
Contour levels for PC1 (a) are every 10$\sigma$ from 3$\sigma$, where $\sigma$ is 3, and those for PC2 (b) are every 3$\sigma$ from 3$\sigma$, where $\sigma$ is 4.
The red circles show the beam size.
(c, d) Channel maps of the first two principal components. Each panel represents the integrated intensity over a velocity range of 1.4 \kms\ except for the last one (the panel of 13.8 \kms), whose lower-end velocity is quoted on the upper-left corner. The panel of 13.8 \kms\ is integrated over a velocity range of 0.7 \kms. The systemic velocity is 8.34 \kms. 
Contour levels for PC1 (c) and PC2 (d) are every 3$\sigma$ from 3$\sigma$, where $\sigma$ is 0.6 and 0.8, respectively.
The cross marks show the continuum peak position. \label{chan_B335}}
\end{minipage}}
\end{figure}

\begin{figure}[h!]
\rotatebox{90}{
\begin{minipage}{\textheight}
\centering

\includegraphics[scale=0.5]{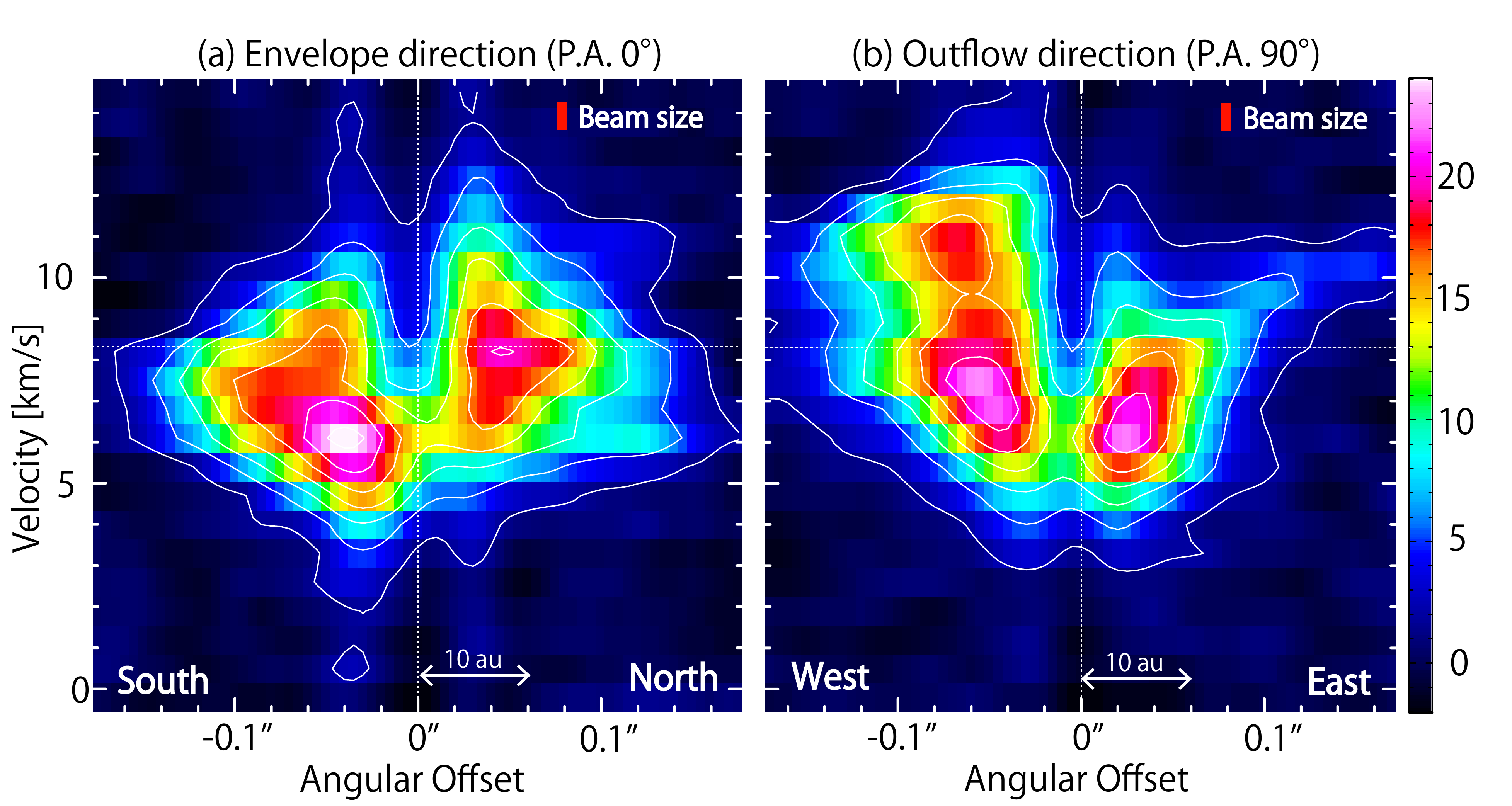}

\caption{Position-velocity diagrams of PC1 along the envelope (a) and outflow (b) directions (Figure \ref{cont_B335}). The horizontal dotted lines represent the systemic velocity of 8.34 \kms. 
Contour levels are every 3$\sigma$ from 3$\sigma$, where $\sigma$ is 0.6.
\label{pvpc1}}
\end{minipage}}
\end{figure}

\begin{figure}[h!]
\centering

\includegraphics[scale=0.7]{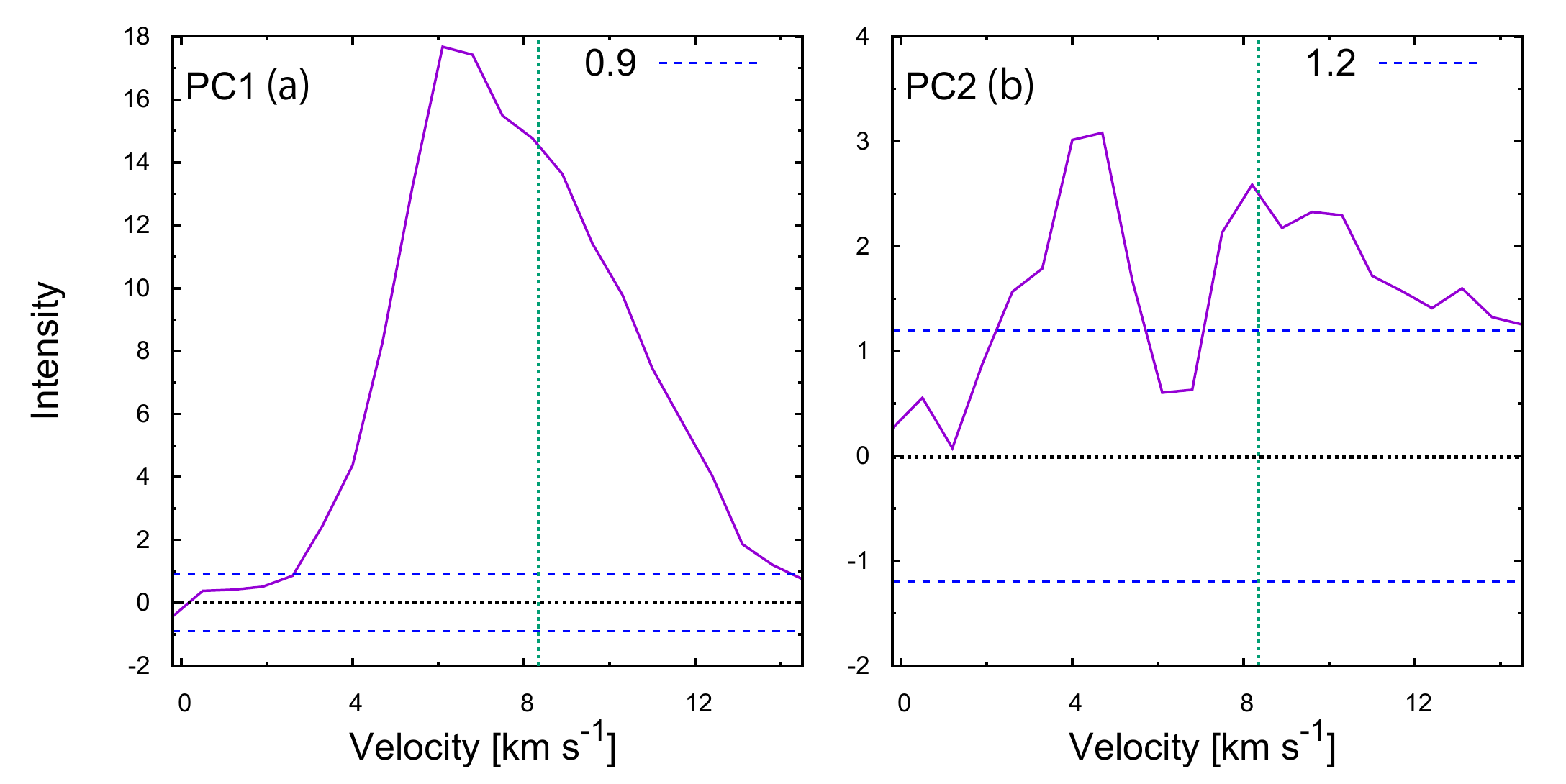}

\caption{Spectral line profiles of the first two principal components in PCA-3D.
The spectra are prepared for a circular region within a diameter of 0.$''$1 centered at the continuum peak.
The horizontal blue dashed lines represent $\pm$3$\sigma$ noise levels, where $\sigma$ for (a) PC1 and (b) PC2 is 0.3 and 0.4, respectively. The horizontal black dotted lines represent the zero-level intensity. The vertical green dotted lines represent the systemic velocity of 8.34 \kms. 
\label{pcspec_B335}}
\end{figure}

\begin{figure}[h!]
\centering

\includegraphics[scale=0.5]{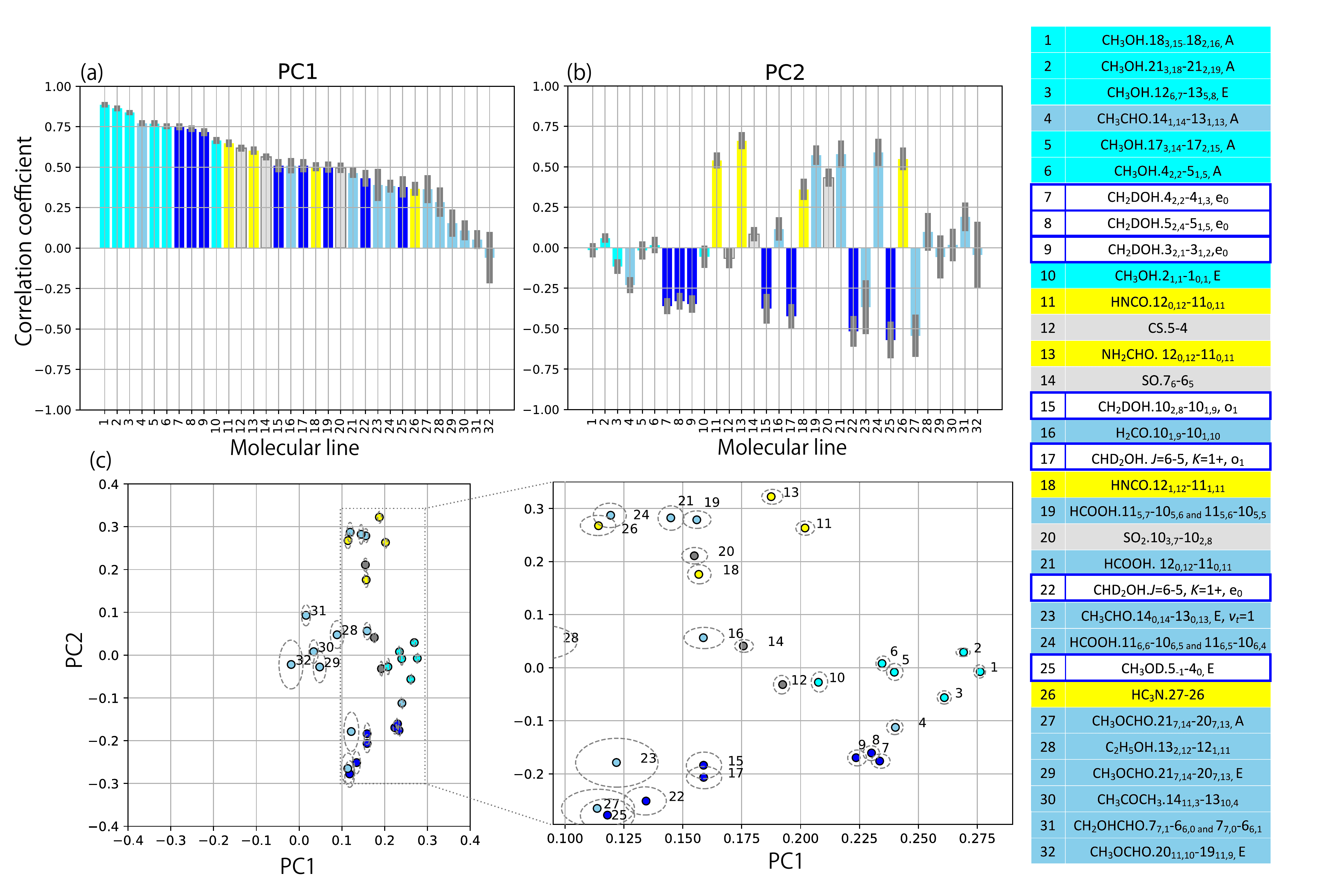}
\caption{(a,b) Correlation coefficients between the first two principal components in PCA-3D and the molecular lines. The uncertainties are shown in grey, which are evaluated by the method reported by \cite{Okoda et al.(2020), Okoda et al.(2021)}.
The numbers represent the molecular lines listed in the attached table.
(c)  Biplots of the contributions for the principal components in PCA-3D for each molecular-line distribution on the PC1-PC2 plane.
The dashed ellipses represent the uncertainties, which are evaluated by the method reported by \cite{Okoda et al.(2020), Okoda et al.(2021)}. The numbers represent the molecular lines listed in the attached table.
Cyan, blue, and light blue plots represent \meta, deuterated \meta\ (\metd, CHD$_2$OH, and CH$_3$OD), and the other oxygen-bearing molecules, respectively. Yellow plots indicate the nitrogen-bearing organic molecules. \label{load_B335}}
\end{figure}

\begin{figure}[h!]
\centering

\includegraphics[scale=0.36]{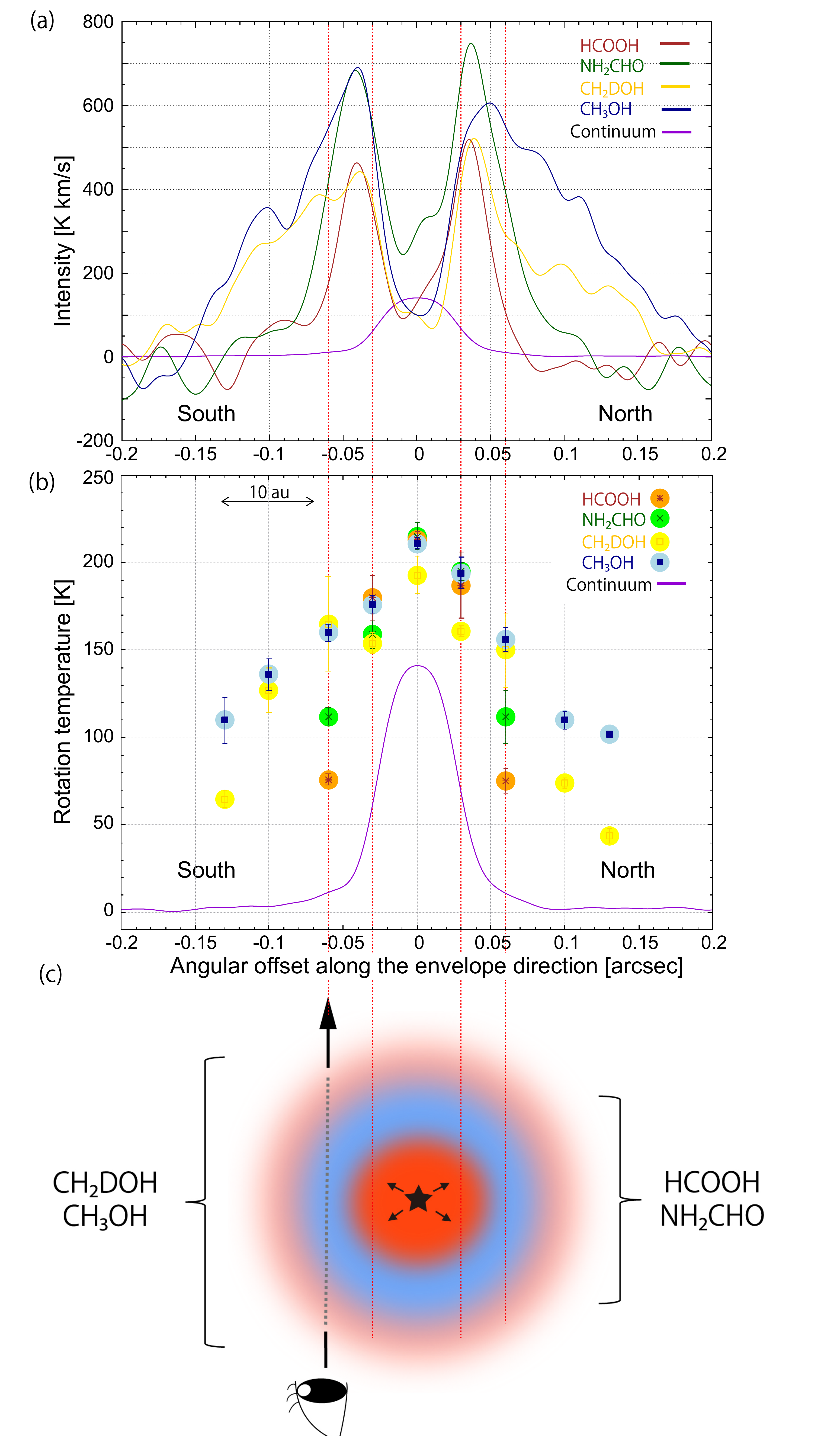}

\caption{(a) Intensity profiles of the HCOOH (12$_{0,12}-11_{0,11}$), \nhhcho\ (12$_{0,12}-11_{0,11}$), \metd\ (4$_{2,2}-4_{1,3}$, e$_0$), \meta\ (4$_{2,2}-5_{1,5}$, A) lines, and the dust continuum emission along the envelope direction (Figure \ref{cont_B335}).
The intensity depression of the molecular lines at the continuum peak is due to the high dust opacity.
(b) Rotation temperature distribution along the envelope direction. The quoted errors are the fitting error of the multi-line analysis (Section \ref{temperature_distribution}). (c) Schematic illustration of the disk/envelope system. The temperature of the region in red is higher than than that in blue. Accretion shock by an infalling gas raises the temperature of the outer part. The vertical red dotted lines just relates three panels. \label{disk_temp}}
\end{figure}

\clearpage

\appendix
\section{PC3 and PC4 in PCA-3D}\label{pc3pc4}
\par
\begin{figure}[h!]
\rotatebox{90}{
\begin{minipage}{\textheight}
\centering

\includegraphics[scale=0.4]{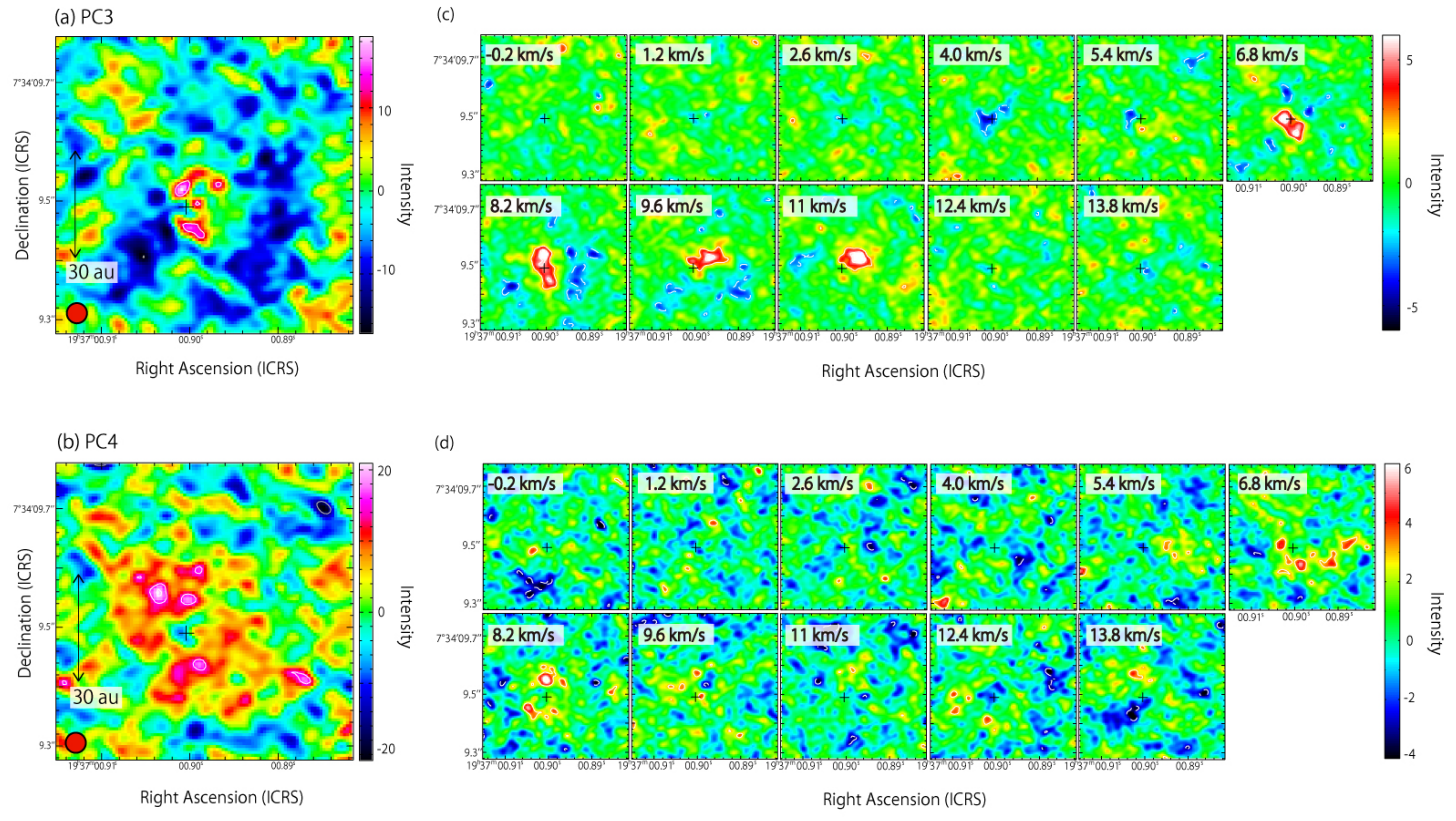}
\caption{(a, b) Moment 0 maps of the third and fourth principal components (PC3 and PC4) in PCA-3D. 
The velocity range of integration is from -0.2 \kms\ to 14.5 \kms.
Contour levels are every 3$\sigma$ from 3$\sigma$, where $\sigma$ is 5.
The red circles show the beam size.
(c, d) Channel maps of the third and fourth principal components. Each panel represents the integrated intensity over a velocity range of 1.4 \kms\ except for the last one (the panel of 13.8 \kms), whose lower-end velocity is quoted on the upper-left corner. The panel of 13.8 \kms\ is integrated over a velocity range of 0.7 \kms. The systemic velocity is 8.34 \kms. 
Contour levels for PC3 (c) and PC4 (d) are every 3$\sigma$ from 3$\sigma$, where $\sigma$ is 0.9 and 1.0, respectively.
The cross marks show the continuum peak position.\label{chan_B335pc3pc4}}
\end{minipage}}
\end{figure}

 The distributions of PC3 and PC4 have the low signal-to-noise ratio in comparison with those of PC1 and PC2. 
The contribution ratios of PC3 and PC4 are 7.9 \% and 5.0 \%, respectively.
Nevertheless, some structures are seen in each velocity channel map, as shown in Figure \ref{chan_B335pc3pc4}.
We here describe the specific features of PC3 and PC4 and the characteristics of the molecular lines extracted by them.
The eigenvectors and eigenvalues for PC3 and PC4 are listed in Table \ref{PCA3D_B335}.
\par
\begin{figure}[h!]
\centering

\includegraphics[scale=0.7]{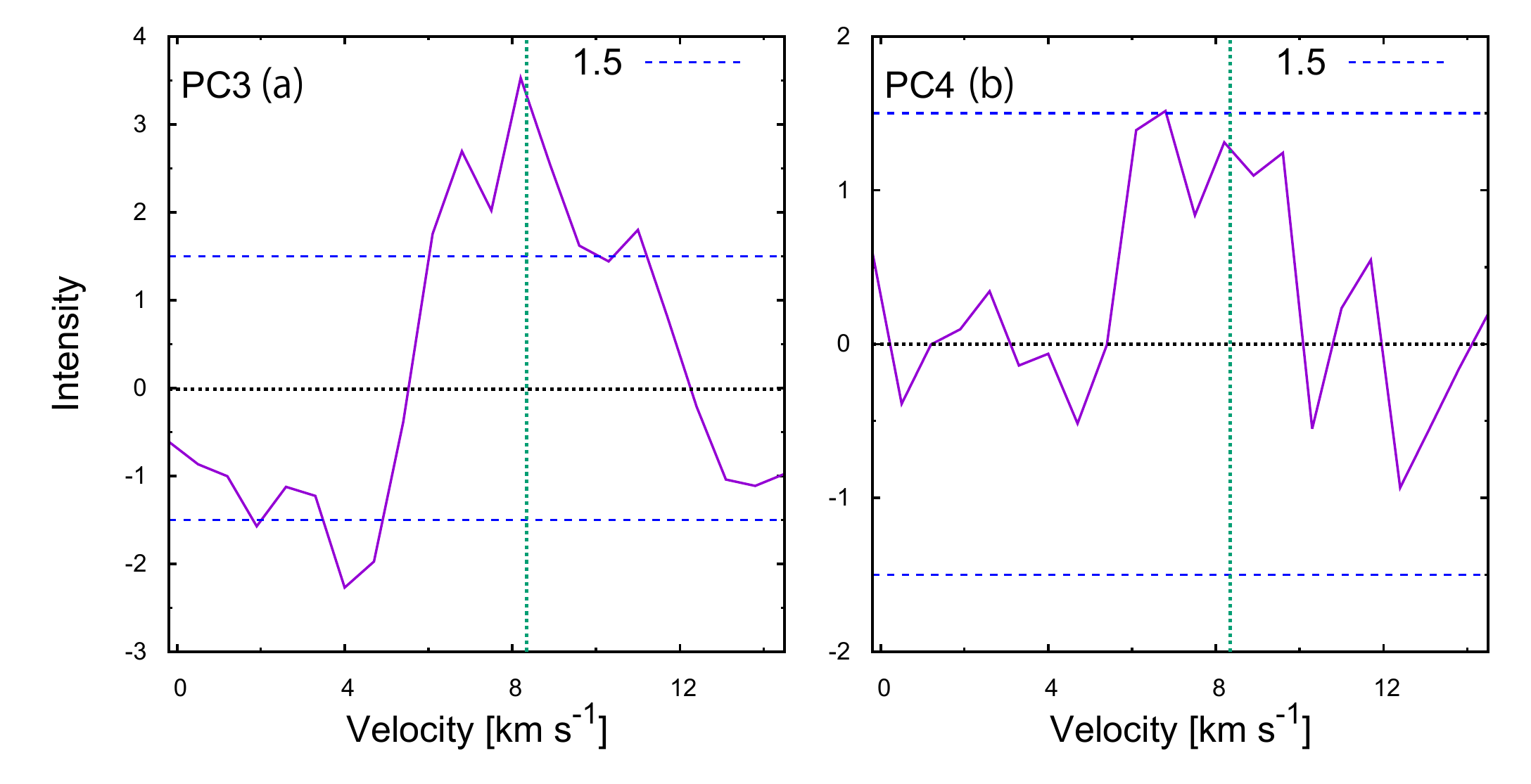}

\caption{Spectral line profiles of the third and fourth principal components in PCA-3D.
The spectra are prepared for a circular region with a diameter of 0.$''$1 centered at the continuum peak.
The horizontal blue dashed lines represent $\pm$3$\sigma$ noise levels, where $\sigma$ is 0.5. The horizontal black dotted lines represent the zero-level intensity. The vertical green dotted lines represent the systemic velocity of 8.34 \kms. \label{pcspec_B335_pc3pc4}}
\end{figure}

 In Figure \ref{chan_B335pc3pc4} (c), PC3 has a positive compact distribution with negative extended clumpy features from 6.8 \kms\ to 11.0 \kms.
As shown in Figures \ref{chan_B335pc3pc4}(a), the intensity of the southern side is negatively brighter than that of the northern side. 
Meanwhile, PC4 shows a weak emission surrounding the continuum peak (Figure \ref{chan_B335pc3pc4}(b)), where the positive emissions can be seen particularly at the panels of 6.8 \kms\ and 8.2 \kms\ (Figures \ref{chan_B335pc3pc4} (d)).
In the spectral line profiles, PC3 has the positive and negative peaks at the redshifted and blueshifted velocities, respectively (Figure \ref{pcspec_B335_pc3pc4}(a)).
On the other hand, the intensity of PC4 toward the continuum peak shows a positive intensity near the systemic velocity (Figure \ref{pcspec_B335_pc3pc4}(b)), although it is lower than 3 $\sigma$.
This is due to its extended and clumpy distribution.
\par
\begin{figure}[h!]
\centering

\includegraphics[scale=0.5]{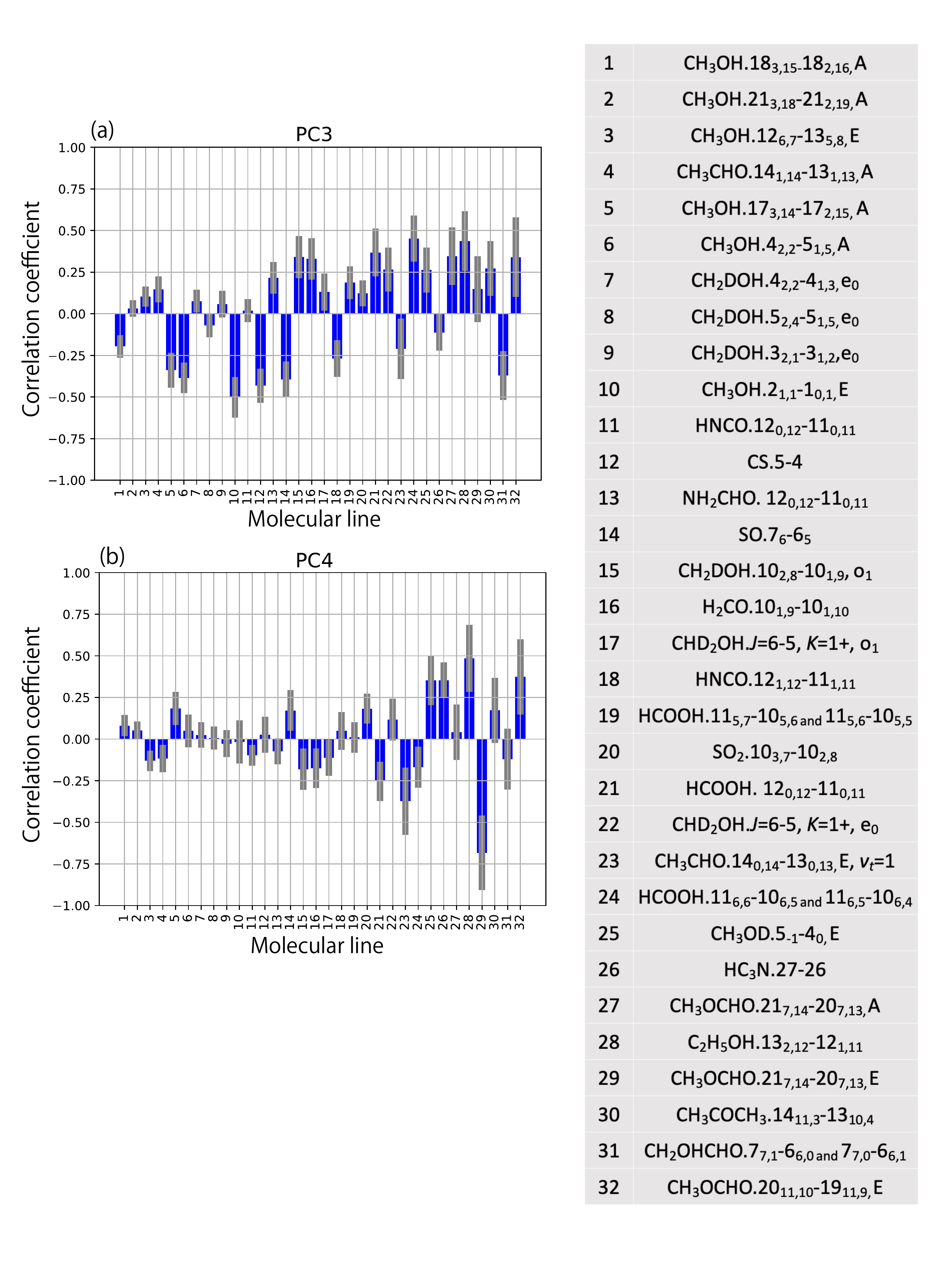}

\caption{(a,b) Correlation coefficients between the principal components and the molecular lines. The uncertainties are shown in grey. The numbers represent the molecular lines listed in the attached table. \label{load_B335pc3pc4}}
\end{figure}

 Figure \ref{load_B335pc3pc4} shows the correlation coefficient between the molecular lines and the principal components.
While all the molecular lines have a correlation smaller than 0.5 for PC3, a few lines have a moderate correlation for PC3 (Figure \ref{load_B335pc3pc4}(a)).
For instance, the CS (\#12) and \meta\ (2$_{1,1}-$1$_{0,1}$, E) (\#10) lines show negative correlation for PC3.
Negative PC3 can represent a more extended distribution than PC1 in the southwestern part.
Hence, it seems reasonable that the low-excitation line (2$_{1,1}-1_{0,1}$, E) (\#10) of \meta\ has a moderate contribution for PC3.
On the other hand, PC4 has negative and positive correlations with CH$_3$OCHO (\#29) and \cchoh\ (\#28), respectively, in particular (Figure \ref{load_B335pc3pc4}(b)).
The moment 0 map of \cchoh (\#28) (Figure \ref{moment_mol_B335}(ab)) is indeed similar to the moment 0 map of PC4.
\par Thus, PC3 and PC4 also contribute to the classification of the data and extract some features of the molecular lines, although their contribution ratios are small.
32 molecular-line data are mostly reproduced by the first four components (PC1-PC4).

\section{Calculation Results from the Multi-line Analysis}\label{app_fitting}
\par We carry out the least-squares analysis on the observed intensities for each molecules at each position by using the method described in Section \ref{LTE}.
Here, examples of the fitting results for the positions of $\pm$0\farcs06 are presented in Table \ref{results_fitting}.
In this calculation, we consider the effect of the optical depth of the dust ($\tau_{\rm dust}$).
The derived $\tau_{\rm dust}$  for each line are summarized in Table \ref{tau_dust}.
The $\tau_{\rm dust}$ values derived from the analysis of different molecules at the continuum peak and the positions of $\pm$0\farcs03 differ by no more than 10 \%. 
Although those at the positions of $\pm$0\farcs06 differ 40 \% at most, the difference is not significant if the uncertainties ($\sigma$) are taken into account.

\newpage

\begin{table}[h!]
\caption{Examples of the Multiline Analysis at the Offsets of $\pm$0\farcs06.\label{results_fitting}}
\scalebox{0.7}{
\begin{tabular}{ccccccccccc}
\hline\hline\noalign{\vskip-.25mm}
Transition & Frequency&$\Delta v$& Intensity & Residual$^{b}$& $\tau_{\rm line}$ &$\Delta v$& Intensity& Residual$^{b}$ & ${\tau_{\rm line}}^{c}$ \\
&(GHz) &(\kms)&(K)&(K)&&(\kms)&(K)&(K)&\\
\hline
\multicolumn{2}{c}{HCOOH}& \multicolumn{4}{c}{-0\farcs06$^a$}& \multicolumn{4}{c}{+0\farcs06$^a$}\\
\hline
11$_{3,8}$$-$10$_{3,7}$ & 248.2744893  &4.9& 54 (8) & -4.2 & 4.6  &5.4& 56 (12)& 0.2 & 3.3\\
11$_{5,7}-10_{5,6}$ and 11$_{5,6}-10_{5,5}$ &247.5140000  &7.1& 57 (10)& 1.9 & 2.8& 6.1&63 (8)& 9.7 & 2.5 \\
11$_{6,6}-10_{6,5}$ and 11$_{6,5}-10_{6,4}$ &247.4462429  &14& 32 (8) & -0.4 & 0.8 &5.4&40 (8)& -6.1 & 1.6 \\
12$_{0,12}-11_{0,11}$ & 262.1034810&5.2& 61 (12)& 2.6 & 6.7&3.5&54 (10)& -3.6 & 8.0 \\
\hline
\multicolumn{2}{c}{\nhhcho}& \multicolumn{4}{c}{-0\farcs06}& \multicolumn{4}{c}{+0\farcs06}  \\
\hline
12$_{0,12}-11_{0,11}$ & 247.3907190&4.9& 97 (9)& 2.8 & 17.5 &3.5& 87 (7)& -8.6 & 24.2 \\
12$_{0,12}-11_{0,11}$, $v_{12}$=1 & 247.3273220&4.0 &33 (8)& -3.6 & 0.5 &5.4& 42 (10)& 12.4 & 0.4 \\
13$_{0,13}-12_{1,12}$ & 244.8542130&3.8& 64 (12)& 2.9 & 1.0 &4.9& 44 (13)& -8.8 & 0.8 \\
\hline
\multicolumn{2}{c}{\metd}& \multicolumn{4}{c}{-0\farcs06}& \multicolumn{4}{c}{+0\farcs06}  \\
\hline
3$_{2,1}-3_{1,2}$, e$_0$ & 247.6257463&4.2&97 (7)& -1.2 & 1.1&4.5& 93 (13)& 7.2 & 1.0 \\
4$_{2,2}-4_{2,3}$, e$_0$ & 244.8411349&4.9&97 (13)& 6.6 & 1.0 &4.2& 83 (15)& -6.5 & 1.1 \\
5$_{2,4}-5_{1,5}$, e$_0$ & 261.6873662&4.0&128 (9)& 4.2 & 1.8 &3.8& 115 (13)& 1.4 & 1.9 \\
10$_{2,8}-10_{1,9}$, o$_1$ & 244.9888456 &4.0&71 (12) & -8.9 & 0.8 &4.9& 58 (8)& -1.9 & 0.6 \\
\hline
\multicolumn{2}{c}{\meta}& \multicolumn{4}{c}{-0\farcs06}& \multicolumn{4}{c}{+0\farcs06}  \\
\hline
2$_{1,1}-1_{0,1}$, E & 261.805675 &5.7& 132 (15)& -10.2 & 6.3 &4.5& 151 (12)& 12.7 & 7.9 \\
4$_{2,2}-5_{1,5}$, A & 247.228587 &5.2&139 (12)& -1.9 & 4.3 &4.9& 131 (13)& -6.2 & 4.5 \\
12$_{6,7}-13_{5,8}$, E & 261.704409 &4.5& 110 (11)& -3.6 & 1.6 &4.2& 109 (9)& -1.5 & 1.6 \\
17$_{3,14}-17_{2,15}$, A & 248.282424 &5.9&140 (11)& -2.6 & 6.7 &4.5& 147 (14)& 8.3 & 8.1 \\
18$_{3,15}-18_{2,16}$, A & 247.610918 &5.2&145 (6)& 2.5 & 6.2 &4.5& 124 (6)& -14.5 & 6.6 \\
21$_{3,18}-21_{2,19}$, A& 245.223019 &4.9&148 (15)& 10.7 & 3.2 &4.2& 142 (15)& 7.9 & 3.4 \\
\hline
\end{tabular}}
\begin{flushleft}
\tablecomments{$^{a}$ The distance from the continuum peak along the envelope (Figure \ref{cont_B335}).
$^{b}$ Residuals of the intensity in the least-square fit (obs.-calc.).
$^{c}$ Optical depth of the line.
Numbers in the parentheses represent the rms noise of the spectrum at each position ($\sigma$).}
\end{flushleft}
\end{table}

\begin{table}[h!]
\caption{\centering{Dust Optical Depth ($\tau_{\rm dust}$) Derived in the Fit$^a$\label{tau_dust}}}
\scalebox{0.7}{
\begin{tabular}{cccccccccc}
\hline\hline
Position &Offset ($''$)& HCOOH &   \nhhcho& \metd &  \meta&   $I_{\rm dust}$ (K) \\
\hline
Continuum peak & 0 & 1.15 $^{+0.12}_{-0.03}$ & 1.12 $^{+0.13}_{-0.01}$ & 1.41 $^{+0.24}_{-0.08}$ & 1.16 $^{+0.10}_{-0.04}$& 141.1 \\
1 & 0.13 & - & - & 0.07 $\pm0.07$ & 0.03 $\pm0.03$ & 2.4 \\
2 & 0.10 & - & - & 0.03 $\pm0.04$ & 0.02 $\pm0.03$ & 1.9 \\
3 & 0.06 & 0.17$^{+0.06}_{-0.04}$ & 0.11$^{+0.04}_{-0.02}$ & 0.08 $^{+0.03}_{-0.01}$  & 0.08 $\pm0.02$ & 11.0 \\
4 & 0.03 & 0.47 $^{+0.09}_{-0.03}$ & 0.45 $^{+0.04}_{-0.02}$  & 0.58 $^{+0.06}_{-0.03}$  & 0.45 $^{+0.05}_{-0.01}$  & 68.1 \\
5 & -0.03 & 0.44 $^{+0.07}_{-0.01}$ & 0.52 $^{+0.07}_{-0.01}$  & 0.54 $^{+0.07}_{-0.02}$  & 0.45 $^{+0.05}_{-0.02}$  & 61.7 \\
6 & -0.06 & 0.18 $^{+0.06}_{-0.05}$ & 0.11 $\pm0.03$ & 0.07 $^{+0.03}_{-0.01}$ & 0.08$\pm0.02$ & 11.5 \\
7 & -0.10 & - & - & 0.03 $\pm0.02$ & 0.03 $\pm0.02$ & 3.4 \\
8 & -0.13 & - & - & 0.05$\pm0.05$ & 0.03 $\pm0.03$ & 2.9 \\
\hline
\end{tabular}}
\begin{flushleft}
\tablecomments{$^a$ The errors are estimated from the derived temperature based on the equation (5).
}
\end{flushleft}
\end{table}

\end{document}